\documentclass[%
reprint,
superscriptaddress,
nofootinbib,
amsmath,
amssymb,
aps,
prd,
floatfix
]{revtex4}
\usepackage{amsmath, amssymb}
\usepackage{listings}
\usepackage{graphicx}
\usepackage{dcolumn}
\usepackage{bm}
\usepackage{float,color}
\usepackage{xcolor}
\usepackage[normalem]{ulem}
\usepackage{tabularx}
\usepackage{mathtools} 
\usepackage{multirow}

\usepackage{scrextend}
\usepackage[a4paper, total={178mm,263mm},left=17mm,right=15mm,top=17mm]{geometry}

\usepackage{scrextend}
\usepackage{hyperref}

\allowdisplaybreaks

\begin{document}

\title{
Parametrized multipolar gravitational waveforms for testing general relativity: Amplitude corrections up to 2PN order
}

\author{Parthapratim Mahapatra}\email{ppmp75@cmi.ac.in}
\affiliation{Chennai Mathematical Institute, Siruseri, 603103, India}
\author{Shilpa Kastha}
\affiliation{Niels Bohr International Academy, Niels Bohr Institute, Blegdamsvej 17, 2100 Copenhagen, Denmark}

\date{\today}

\begin{abstract}
A parametrized multipolar gravitational wave phasing within multipolar post-Minkowskian and post-Newtonian formalism was developed in earlier works [S. Kastha {\it et al.}, PRD {\bf 98}, 124033 (2018) and PRD {\bf 100}, 044007 (2019)]. This facilitates the model-agnostic tests for the multipolar structure of compact binaries using gravitational wave observations.
In this paper, we derive a parametrized multipolar {\emph {amplitude}} of the gravitational wave signal in terms of mass and current-type radiative multipole moments within the post-Newtonian approximation to general relativity.
We assume the compact binary to be moving in quasicircular orbits, with component spins (anti-) aligned with respect to the binary's orbital angular momentum. We report a closed-form expression for the parametrized multipolar amplitude of the waveform at second post-Newtonian order both in time and frequency domains. This includes the contribution from the leading five mass-type and the leading four current-type radiative moments.
This framework of constructing a parametrized waveform accomplishes a generic parametrization of {\emph {both}} gravitational wave \emph{phase} and \emph{amplitude} with the \emph{same set of phenomenological parameters}. Hence, it should significantly enhance the precision of the multipole tests in the context of present and future gravitational wave detectors.
\end{abstract}

\maketitle

\section{INTRODUCTION}\label{sec:intro}
Gravitational waves (GWs) carry information from regions of spacetime where gravity is ``strong", and therefore are genuine probes of the predictions of general relativity (GR). Advanced LIGO~\cite{Aasi_2015} and advanced Virgo~\cite{Acernese_2015} have now detected nearly a hundred compact binary mergers~\cite{GW150914,GWTC1,GWTC2,GWTC3, Nitz:2021zwj,Nitz:2021uxj,Venumadhav:2019lyq}, a subset of which has been used to carry out various kinds of tests of GR~\cite{TGR-GWTC-1,TGR-GWTC-2,TGR-GWTC-3}. These tests have found the compact binary signals to be consistent with GR~\cite{TGR-GWTC-1,TGR-GWTC-2,TGR-GWTC-3} so far. 
However, as the GW detectors improve their sensitivity~\cite{CEScience,CEScience19,ETScience10}, GR can be tested to unprecedented precision using future GW experiments. Further, space-based detector LISA~\cite{LISAScience17}, expected to be launched in 2032, can facilitate novel probes of strong-gravity effects.
Besides attempts to directly probe specific alternatives to GR~\cite{Ma:2023sok}, theory-agnostic parametrized tests are very efficient tools to perform such tests.
Theory-agnostic tests of GR search for generic consistency between GR waveforms and the observed data without invoking a specific theory of gravity. Examples for this class of tests include parametrized tests of post-Newtonian (PN) theory~\cite{AIQS06a, AIQS06b, MAIS10, TIGER, LiEtal2012, PhysRevD.97.044033}, parametrized post-Einstein formalism~\cite{YunesPretorius09}, bounding the mass of the graviton and dispersion of GWs~\cite{Will98,MYW11}, inspiral-merger-ringdown consistency tests~\cite{Ghosh2016, Kastha:2021chr} and tests of multipolar gravitational radiation~\cite{Dhanpal:2018ufk,Islam:2019dmk,Kastha:2018bcr,Kastha:2019,Capano:2020dix,Puecher:2022sfm}. 

Within the multipolar post-Minkowskian and post-Newtonian (MPM-PN) approximation~\cite{Th80,BD84,BD86,B87,BD88,BD89,BD92,B95,BDI95,BIJ02,DJSdim,BDEI04} of GR, the gravitational field of an inspiralling compact binary can be modeled in terms of mass- and current-type multipole moments. These multipole moments and their interactions during the dynamical evolution of the binary lead to various physical effects arising at different PN orders in the gravitational waveform. More specifically, in this formalism~\cite{Th80,BD84,BD86,B87,BD88,BD89,BD92,B95,BDI95,BIJ02,DJSdim,BDEI04}, the radiation content in the far zone of the source is encoded in the mass- and current-type radiative multipole moments $\{U_L,V_L\}$ which can be related to the stress-energy tensor of the compact binary. Hence, $\{U_L,V_L\}$ carry all the relevant information about the gravitational fields arising from the compact binary evolution, and their contributions are manifested through the GW phase and amplitude.

In an alternative theory of gravity, one or more multipole moments may differ from those predicted by GR~\cite{Endlich:2017tqa,Compere:2017wrj,Bernard:2018hta,Julie:2019sab,Shiralilou:2021mfl,Battista:2021rlh,Bernard:2022noq,Julie:2022qux,Diedrichs:2023foj}. Recently, in a series of two papers~\cite{Kastha:2018bcr, Kastha:2019}, 
a unique theory-agnostic test was proposed to probe the PN multipolar structure of the gravitational field emitted by an inspiralling compact binary in quasicircular orbits. They discuss how well these radiative multipole moments can be constrained using GW observations. In paper I~\cite{Kastha:2018bcr}, to capture the generic deviations, the parametric deformations are introduced at the level of mass-type ($U_L$) and current-type ($V_L$) radiative multipole moments as follows:
\begin{align}\label{eq:muepsilon}
&U_L\rightarrow \mu_l\,U_L^{\rm GR}\, ,\nonumber \\ 
&V_L \rightarrow \epsilon_l \, V_L^{\rm GR} \, ,
\end{align}
with $\mu_l=1+\delta U_L/U_L^{\rm GR}$ and  $\epsilon_l=1+\delta V_L/V_L^{\rm GR}$. Here, $\delta U_L$ and $\delta V_L$ are the deviations of the mass and the current-type radiative multipole moments from their GR counterparts. With this parametrization the corresponding gravitational wave phasing is recomputed {upto 3.5PN order}. By construction, these phenomenological multipole parameters $\mu_l$, $\epsilon_l$ are identically equal to unity in GR. With the above parametrization, we can explicitly keep track of the contributions from various radiative multipole moments to the GW phasing. Hence, we can perform tests of the multipolar structure of the PN approximation to GR. 

Following the similar strategy, a subsequent paper~\cite{Kastha:2019} extended this test to the case of nonprecessing spinning binaries.
In the computation of gravitational waveform, the orbital binding energy is another important constituent. In the paper II~\cite{Kastha:2019}, the deviations at each PN order of the conserved orbital energy are also considered and hence are able to constrain the orbital energy from the observations. However, both of these papers accounted for only the leading quadrupolar contribution to the amplitude of the gravitational waveform.

The leading-order gravitational wave emission is quadrupolar, according to GR. However, compact binary mergers with unequal mass ratios produce gravitational radiation that have non-negligible power in many harmonics of the fundamental mode. In case of GW parameter estimation with restricted waveforms, the source properties are extracted only from the phasing.
It has been shown in various previous studies that the amplitude-corrected waveforms, which encode a wealth of additional information by including the contribution from higher harmonics, lead to dramatic improvements in inferring the source properties~\cite{Sintes:1999cg, Chris06, ChrisAnand06, VanDenBroeck:2006ar}. The observed strengths of the higher harmonics depend on the inclination angle $\iota$ and the fractional mass difference, $\delta = \frac{m_{1} -m_{2}}{m_{1} + m_{2}}$, with $m_{1}$, $m_{2}$ ($m_{1}>m_{2}$) being the component masses~\cite{BFIS08, VanDenBroeck:2006qu, VanDenBroeck:2006ar, ABFO08, MKAF2016, IMRPhenomHM, Roy:2019phx, Khan:2019kot, Cotesta:2018fcv}. Strong evidence for the presence of higher-order multipole moments is also found in the recent detection of GW signals from GW190412 and GW190814 by LIGO/Virgo interferometers \cite{GW190412, GW190814}. For both the systems, the component masses are found to be asymmetric, and the posterior distribution of the inclination angle is suggestive of the systems being not face-on/face-off ($\iota=0$/$\iota=180^{\circ}$).

Recently, several efforts have been devoted to devise theory-agnostic tests of GR in a parametric way that directly probes the higher harmonic structure for binary black hole (BBH) coalescences 
\cite{Dhanpal:2018ufk, Kastha:2018bcr, Kastha:2019, Islam:2019dmk, Capano:2020dix, Puecher:2022sfm, Mezzasoma:2022pjb}. In Refs.~\cite{Dhanpal:2018ufk, Islam:2019dmk}, a “no-hair” test is developed for BBHs by examining the consistency of the source parameters inferred from just the quadrupole (leading-order) mode and the higher-order modes separately. Using the data of the two GW events, GW190412 and GW190814, Ref.~\cite{Capano:2020dix}, tested the consistency between the dominant and subdominant modes and find good agreement with GR. More recently, Ref.~\cite{Puecher:2022sfm} developed a test that investigates the consistency of the (3, 3) and (2, 1) modes amplitudes with GR prediction. Applying the test to GW190412 and GW190814, they also find no evidence for violation of GR. Furthermore, Ref.~\cite{Mahapatra:2023hqq} showed that the post-Newtonian structure of octupolar radiation from these two GW events, also considering the intriguing effect of octupolar tails, is consistent with the predictions of GR. Parallelly, in order to include the effects from the second, third, and the fourth harmonics, Ref.~\cite{Mezzasoma:2022pjb} has provided a 1PN amplitude corrected waveform for nonspinning binaries in circular orbits to perform the test of GR within the parametrized post-Einsteinian framework \cite{Yunes:2009ke}.

Instead of parametrizing the GW phase and amplitude independently, a consistent simultaneous parametrization of both the \emph{phase} and the \emph{amplitude} of the GW enables us to search for GR deviations with a relatively \emph{smaller} set of phenomenological parameters.
Furthermore, by introducing multipole parameters into the amplitude, one expects the constraints on different multipole moments to be more stringent and should increase our ability to detect a GR violation, if present.
This motivates us to compute the parametrized multipolar amplitude corrections at higher PN orders. In this paper, we extend the computation of parametrized multipolar waveform, developed in \cite{Kastha:2018bcr, Kastha:2019}, to higher harmonics. We provide a ready-to-use PN waveform in both the time and the frequency domains for compact binaries in quasicircular orbits with the spins of the component masses being (anti-) aligned with respect to the orbital angular momentum at 2PN. This modified waveform amplitude includes the contributions from the first six harmonics of the orbital phase. 

The reported amplitude-corrected waveform model forms the basis for performing the multipolar tests of compact binary evolution in GR. It helps in deducing the constraints on the parameter space of the theories of gravity which predicts a different multipole structure than GR~\cite{Endlich:2017tqa,Compere:2017wrj,Bernard:2018hta,Julie:2019sab,Shiralilou:2021mfl,Battista:2021rlh,Bernard:2022noq,Julie:2022qux,Diedrichs:2023foj}. Moreover, the complete expression of this waveform can be used to map the constraints on the multipole parameters to the other existing parameters in various parametrized tests of GR~\cite{Mahapatra:2023uwd}. This also serves as the starting point of constructing a complete parametrized phenomenological inspiral-merger-ringdown waveform model for testing the multipole structure in GR using the GW data which we keep for future explorations.

The remainder of the paper is organized as follows. In Sec.~\ref{sec:TD-waveform}, we show the form of the time domain multipolar waveform, specify our notations, and report the expressions for time domain parametrized multipolar polarizations. Section~\ref{sec:FD-waveform} presents the structure of the frequency domain parametrized multipolar polarizations and their closed-form expressions. Finally, in Sec.~\ref{sec:conclu}, we summarize our findings and discuss the future directions.

\section{parametrized MULTIPOLAR GRAVITATIONAL WAVE AMPLITUDE in time domain}\label{sec:TD-waveform}
The two-body dynamics and the emission of gravitational waves in the context of compact binary mergers within GR can be solved perturbatively using the MPM-PN formalism in the adiabatic regime, where the orbital time scale is much smaller than the radiation reaction time scale~\cite{Th80,BD84,BD86,B87,BD88,BD89,JunkS92,BD92,B95,BDI95,BIJ02,DJSdim,BDEI04}. This MPM-PN formalism can be portrayed as a smooth blend of post-Minkowskian approximation (expansion of the gravitational metric in powers of G, Newton’s gravitational constant, valid throughout the spacetime for weakly gravitating sources), PN expansions (an expansion in 1/c, inverse of the speed of light, that is valid for slowly moving and weakly gravitating sources and applicable in the near zone of the source) and the multipole expansion technique of the gravitational field valid over the entire spacetime region exterior to the source (see Ref.~\cite{Bliving} for a review). Within this formalism, the gravitational wave strain from compact binary systems can be uniquely decomposed in terms of the radiative multipole moments \cite{Th80,BDI95}, 
\begin{eqnarray}
 h^{TT}_{ij} &=& \frac{4G}{c^2 D} {\cal P}_{ijkm} ({\bf N})
 \sum^\infty_{\ell =2} \frac{1}{c^\ell \ell !} \biggl\{ N_{L-2} U_{kmL-2}(T_R) - \frac{2\ell}{(\ell +1)c} N_{aL-2} \varepsilon_{ab(k} V_{m)bL-2}  (T_R)\biggr\} + \mathcal{O} \left( \frac{1}{{D}^2}\right), \label{eq:Strain}
\end{eqnarray}
where the superscript TT stands for the transverse-traceless projection with ${\cal P}_{ijkm}$ being the usual TT algebraic projector.
Here {\bf N} is the direction of GW propagation (i.e., the radial direction from the radiating source to the observer), $T_R$ is the retarded time, $D$ is the proper distance to the detector, $\varepsilon_{ijk}$ is the antisymmetric Levi-Civita tensor. The radiative multipole moments $U_L$ and $V_L$ (with $L = ij$··· a spatial multi-index composed of $\ell$ ($\ell \ge 2$) indices) are the set of symmetric trace-free (STF) three-dimensional Cartesian tensors of rank $\ell$. These radiative moments $U_L$ and $V_L$ can be rewritten as the nonlinear functional of the “canonical” source multipole moments \{$M_L$, $S_L$\} and eventually is mapped onto the source multipole moments \{$I_L$, $J_L$,...\}~~\cite{BD84,BD86,B87,BD88,BD89,BD92,BDI95}. These source multipole moments of the relativistic mass and current densities are then expressed as functionals of the stress-energy pseudotensor $\tau_{\mu \nu}$ of the source and its gravitational fields~\cite{Bliving,BDI95,ABIQ04}.
The relations between radiative and source moments consist of various nonlinear multipolar interactions such as tails~\cite{BD87,BS93,BDI95}, tails of tails~\cite{B98tail}, tail square~\cite{B98quad}, memory~\cite{Chr91,Th92,ABIQ04,Favata08}, etc., as the gravitational waves propagate from the source to the detector (see Ref.~\cite{Bliving} also).

To compute the parametrized multipolar waveform with the amplitude correction at 2PN order, we closely follow the prescription given in Refs.~\cite{BDI95,WWi96,BIWW96,ABIQ04,ABFO08,MKAF2016}. Using the non-GR multipolar deformations mentioned in Eq.~(\ref{eq:muepsilon})~\cite{Kastha:2018bcr,Kastha:2019}, the parametrized multipolar waveform in Eq.~(\ref{eq:Strain}) at 2PN reads,
\begin{eqnarray}
 h^{TT}_{ij} = \frac{2G}{c^4 D} {\cal P}_{ijkm} \biggl\{ \mu_{2} U_{km} 
 &+& \frac{1}{c} \left[ \frac{1}{3} \mu_{3} N_a  U_{kma} + \frac{4}{3} \epsilon_{2}  \varepsilon_{ab(k} V_{m)a} N_b \right] \nonumber \\
  &+& \frac{1}{c^2} \left[ \frac{1}{12} \mu_{4} N_{ab} U_{kmab} + \frac{1}{2} \epsilon_{3}  \varepsilon_{ab(k} V_{m)ac} N_{bc} \right] \nonumber \\
  &+& \frac{1}{c^3} \left[ \frac{1}{60} \mu_{5} N_{abc} U_{kmabc} + \frac{2}{15}
  \epsilon_{4} \varepsilon_{ab(k} V_{m)acd} N_{bcd} \right] \nonumber \\
  &+& \frac{1}{c^4} \left[ \frac{1}{360} \mu_{6} N_{abcd}  U_{kmabcd} + \frac{1}{36}
  \epsilon_{5} \varepsilon_{ab(k} V_{m)acde} N_{bcde} \right] \nonumber \\
  &+& \mathcal{O}(\frac{1}{c^5})
   \biggr\}\ .
   \label{eq:MultipolarStrain}
\end{eqnarray}

This can be reexpressed using previously reported closed-form expressions of the radiative multipole in terms of mass- and current-type source multipole moments \{$I_L$, $J_L$,..\}~\cite{Bliving,BDI95,ABIQ04}. We finally compute the multipolar waveform in terms of the binary source parameters using the 2PN corrected expressions of \{$I_L$, $J_L, ...$\} in terms of the binary parameters and the two-body equations of motion. We do not quote the detailed expression of the source multipole moments here and point to Refs.~\cite{BDI95, BIWW96, ABIQ04} for nonspinning (NS) effects, Refs.~\cite{BBuF06, BBF2011, BMB2013} for spin-orbit (SO) terms and Refs.~\cite{KWWi93, K95, WWi96, Racine2008, BFMP2015} for the contributions from the spin-spin (SS) interactions. For all the computations reported here, we follow the notations and definitions of Ref.~\cite{ABFO08}. 

For the rest of the article, we use the following symbols for various parameters characterizing the compact binary system: the component masses are $m_1$, $m_{2}$ (with $m_1>m_2$), and the individual spin angular momenta are $\mathbf{S}_1$, $\mathbf{S}_2$. We define the total mass to be $M$ and the symmetric mass ratio to be $\nu$.
The normalized dimensionless individual spin vectors are denoted by $\boldsymbol{\chi}_b$ (b = 1,2), whereas the symmetric combination is denoted as $\boldsymbol{\chi}_s$ and antisymmetric combination by $\boldsymbol{\chi}_a$. The explicit mathematical expressions for these quantities are the following:
\begin{eqnarray}
M & = & m_1 + m_2\,, \nonumber \\
\delta & = & \frac{m_1-m_2}{m_1+m_2}\,, \nonumber\\
\nu  &=&  \frac{m_1\,m_2}{M^2}= \frac{1}{4}(1 - \delta^2), \label{massspinparams}\\
\boldsymbol{\chi}_b & = &
\frac{\mathbf{S}_b}{G m_b^2 / c}\,, \quad (b = 1,2)\, \nonumber\\
\boldsymbol{\chi}_s & = & \frac{1}{2}\left(\boldsymbol{\chi}_1 +
\boldsymbol{\chi}_2 \right)\,, \nonumber\\ 
\boldsymbol{\chi}_a & = &
\frac{1}{2}\left(\boldsymbol{\chi}_1 - \boldsymbol{\chi}_2
\right) \nonumber. 
\end{eqnarray} 
It is evident from the above relations that the symmetric mass ratio satisfies $0 < \nu \leq 1/4$, the fractional mass difference obeys $0 < \delta < 1$, and the magnitude of $\boldsymbol{\chi}_b$, $\boldsymbol{\chi}_s$, and $\boldsymbol{\chi}_a$ are bounded between 0 and 1.

We now provide the two polarization states of gravitational wave strain tensor $h_+$ and $h_\times$,  needed for all the practical purposes. These are defined by
\begin{eqnarray}
h_+ = {1 \over 2} (P_i P_j - Q_i Q_j) h_{ij}^{\rm TT} \,, \nonumber
\\ h_\times
= {1 \over 2} (P_i Q_j + Q_i P_j) h_{ij}^{\rm TT} \,,
\end{eqnarray}
where ${\bf P}$ and ${\bf Q}$ denote two unit polarization vectors, along with the direction ${\bf N}$, forming an orthonormal right-handed triad. In the case of nonprecessing circular binary systems the standard convention is to choose ${\bf P}$ lying along the intersection of the orbital plane with the plane of the sky in the direction of the “ascending node” and ${\bf Q} = {\bf N} \times {\bf P}$ (see Sec. VII of \cite{WWi96}). According to this convention the unit radial vector joining the two bodies is given by ${\bf n}= {\bf P} \cos\Phi + ({\bf Q} \cos i + {\bf N} \sin i) \sin \Phi$, where $i$ denotes the orbit's inclination angle and $\Phi$ is the orbital phase, namely the angle between the ascending node and the direction of body one (see Fig. 7 of \cite{WWi96}). The unit direction of the orbital velocity is given by $\bm{\lambda}= -{\bf P} \sin \Phi + ({\bf Q} \cos i + {\bf N} \sin i) \cos \Phi$ and the unit direction along the Newtonian orbital angular momentum is given by $\mathbf{\hat{L}}_{\rm N}\equiv {\bf n} \times \bm{\lambda} = - {\bf Q} \sin i + {\bf N} \cos i$.

Following Refs.~\cite{WWi96, BIWW96, ABIQ04, ABFO08} we obtain the time domain parametrized multipolar waveform polarizations for nonprecessing compact binary moving in quasicircular orbits. Similar to the earlier works \cite{WWi96, BIWW96, ABIQ04, ABFO08}, these polarizations are expressed in terms of the gauge-invariant PN parameter $v=(G M \omega)^{1/3}$, where $\omega$ is the orbital frequency of the circular orbit accurate up to 2PN order. For convenience, we first quote the PN schematic structure of the waveform polarizations here,
\begin{eqnarray}
h_{+,\times}\ = \
\frac{2\, G \, M\,\nu\,v^2}{c^4 \, D_L}\,\Bigg[\xi_{+,\times}^{(0,{\rm NS})}+ \frac{v}{c} \, \xi_{+,\times}^{(1/2,{\rm NS})} +
\frac{v^2}{c^2} \,  \Big(\xi_{+,\times}^{(1,{\rm NS})} + \xi_{+,\times}^{(1,{\rm SO})}\Big) +
\frac{v^3}{c^3} \, \Big(\xi_{+,\times}^{(3/2,{\rm NS})} + \xi_{+,\times}^{(3/2,{\rm SO})}\Big)\, \nonumber \\
+ \frac{v^4}{c^4} \, \Big(\xi_{+,\times}^{(2,{\rm NS})} + \xi_{+,\times}^{(2,{\rm SO})} + \xi_{+,\times}^{(2,{\rm SS})}\Big) 
\Bigg]\, .
\label{eq:pol-schematic}
\end{eqnarray}
In the above equation $\xi_{+,\times}^{(n/2,{\rm NS})}$, $\xi_{+,\times}^{(n/2,{\rm SO})}$, and $\xi_{+,\times}^{(n/2,{\rm SS})}$  (n/2 is the PN order with $n=0,1,2\ldots$) indicates the nonspinning, the spin-orbit (linear-in-spins), and the spin-spin (quadratic-in-spin) contributions respectively at the n/2-th PN order. To incorporate the effects of cosmological expansion, we have replaced the proper distance $D$ by $D_{L}$, where $D_{L}$ is the luminosity distance (see Chapter 4.1.4 of \cite{maggiore-GWvol1} and Appendix C of Ref.~\cite{Favata:2021vhw} for full justification). To be noted here that, in Eq.~(\ref{eq:pol-schematic}), the total mass $M$ is the “detector frame” or the “observer frame” total mass of the binary. We now provide the explicit expression for all the components of Eq.~(\ref{eq:pol-schematic}), 

\begin{subequations}
\label{eq:HpTD}
\begin{align}
\xi_{+}^{(0,{\rm NS})} &= -\left(1+{c_i}^2\right) \mu_2 \cos 2 \Psi \, , \\
\xi_{+}^{(1/2,{\rm NS})} &= - s_{i} \, \delta  \Bigg[ \cos \Psi \Big( -\frac{\mu_{3}}{24}+\frac{2 \epsilon_{2}}{3} + \frac{\mu_{3}}{8} c_{i}^2 \Big) - \cos 3 \Psi \, \Big( \frac{9 \mu_{3}}{8} \left( 1 + c_{i}^2 \right) \Big)  \Bigg] \, , \\
\xi_{+}^{(1,{\rm NS})} &=  \cos 2 \Psi  \Bigg[ \frac{107 \mu_{2}}{42}-\frac{\mu_ {4}}{21}+\frac{2 \epsilon_{3}}{3} + \Big( \frac{107 \mu_{2}}{42}+\frac{2 \mu_{4}}{7}-\frac{4 \epsilon_{3}}{3} \Big) c_{i}^{2} -\frac{\mu_{4}}{3} c_{i}^{4} \nonumber \\ 
&\quad+ \nu \bigg( -\frac{55 \mu_{2}}{42}+\frac{\mu_{4}}{7}- 2 \epsilon_{3} + \Big( -\frac{55 \mu_{2}}{42}-\frac{6 \mu_{4}}{7}+4 \epsilon_{3}   \Big) c_{i}^{2} + \mu_{4} c_{i}^{4} \bigg)  \Bigg] \nonumber \\
&\quad - \cos 4 \Psi \Bigg[ \frac{4}{3} {s_{i}}^2 \left(1+{c_{i}}^2\right) \left( 1-3 \nu \right)  \mu_{4}  \Bigg] \, ,\\
\xi_{+}^{(1,{\rm SO})} &= s_{i} \cos \Psi  \Bigg[    \left(\boldsymbol{\chi}_a\cdot\mathbf{\hat{L}}_{\rm N} + \delta \, \boldsymbol{\chi}_s\cdot\mathbf{\hat{L}}_{\rm N}\right) \epsilon_{2}  \Bigg]  \, ,\\
\xi_{+}^{(3/2,{\rm NS})} &= s_{i} \, \delta \cos \Psi \Bigg[ -\frac{\mu_{3}}{9}-\frac{\mu_{5}}{2880}+\frac{17 \epsilon_{2}}{42}+\frac{\epsilon_{4}}{280} + \Big(  \frac{\mu_{3}}{3}+\frac{\mu_{5}}{240}-\frac{\epsilon_{4}}{40} \Big) c_{i}^{2} -\frac{\mu _{5}}{192} c_{i}^{4}  \nonumber \\
&\quad + \nu \bigg( -\frac{\mu_{3}}{36}+\frac{\mu_{5}}{1440}-\frac{10 \epsilon_{2}}{21}-\frac{\epsilon_{4}}{140} + \Big( \frac{\mu_{3}}{12}-\frac{\mu_{5}}{120}+\frac{\epsilon_{4}}{20} \Big) c_{i}
^{2} + \frac{\mu_{5}}{96} c_{i}^{4} \bigg) 
 \Bigg] \nonumber \\
&\quad + \cos 2 \Psi \Big[ -2 \pi  \left(1+c_{i}^2\right) \mu _{2}  \Big] \nonumber \\
&\quad + s_{i} \, \delta \cos 3 \Psi  \Bigg[ -\frac{9 \mu_{3}}{2}+\frac{27 \mu_{5}}{640}-\frac{27 \epsilon_{4}}{40} + \Big( -\frac{9 \mu_{3}}{2}-\frac{27 \mu_{5}}{80}+\frac{81 \epsilon_{4}}{40} \Big) c_{i}^{2} + \frac{81 \mu_{5}}{128} c_{i}^{4} \nonumber \\
&\quad + \nu \bigg( \frac{9 \mu_{3}}{4}-\frac{27 \mu_{5}}{320}+\frac{27 \epsilon_{4}}{20} + \Big( \frac{9 \mu_{3}}{4}+\frac{27 \mu_{5}}{40}-\frac{81 \epsilon_{4}}{20}\Big) c_{i}^{2} -\frac{81 \mu_{5}}{64}  c_{i}^{4} \bigg) \Bigg] \nonumber \\
&\quad + s_{i} \, \delta \cos 5 \Psi  \Bigg[ \frac{625}{384} s_{i}^{2}  \left(1+c_{i}^{2}\right)  \left(1 - 2 \nu \right) \mu_{5} \Bigg] \, ,\\
\xi_{+}^{(3/2,{\rm SO})} &= \cos 2 \Psi \Bigg[ \frac{4}{3} \bigg( \big(1 + c_{i}^{2}\big)\left(\boldsymbol{\chi}_s\cdot\mathbf{\hat{L}}_{\rm N} 
+ \delta\,\boldsymbol{\chi}_a\cdot\mathbf{\hat{L}}_{\rm N}\right) \mu_{2}  \nonumber \\
&\quad + \nu \Big( -\left( 1+c_{i}^{2} \right) \mu _{2} + 2 \left(1-2 c_{i}^{2}\right) \epsilon_{3} \Big) \boldsymbol{\chi}_s\cdot\mathbf{\hat{L}}_{\rm N}  \bigg) \Bigg] \, ,\\
\xi_{+}^{(2,{\rm NS})} &= \pi  s_{i} \, \delta \cos \Psi \Bigg[ \frac{\mu_{3}}{24}-\frac{2 \epsilon_{2}}{3} -\frac{ \mu_{3}}{8} c_{i}^{2} \Bigg] \nonumber \\
&\quad +  \cos 2 \Psi \Bigg[  \frac{2173 \mu_{2}}{1512}+\frac{437 \mu_{4}}{2310}+\frac{17 \mu_{6}}{11880}-\frac{193 \epsilon_{3}}{135}-\frac{2 \epsilon_{5}}{135} \nonumber \\ 
&\quad+ \Big( \frac{2173 \mu_{2}}{1512}-\frac{437 \mu_{4}}{385} - \frac{289 \mu_{6}}{11880}  +\frac{386 \epsilon_{3}}{135}+\frac{22 \epsilon _{5}}{135} \Big) c_{i}^{2} \nonumber \\ 
&\quad+ \Big( \frac{437 \mu_{4}}{330}+\frac{49 \mu_{6}}{792}-\frac{8 \epsilon_{5}}{45} \Big) c_{i}^{4} -\frac{\mu_{6}}{24} c_{i}^{6} \nonumber \\
&\quad+ \nu \bigg(  \frac{1069 \mu_{2}}{216}-\frac{115 \mu_{4}}{198} -\frac{17 \mu_{6}}{2376} +\frac{145 \epsilon _{3}}{27}+\frac{2 \epsilon_{5}}{27} \nonumber \\
&\quad+ \Big( \frac{1069 \mu_{2}}{216}+\frac{115 \mu_{4}}{33}+\frac{289 \mu_{6}}{2376}-\frac{290 \epsilon_{3}}{27}-\frac{22 \epsilon_{5}}{27} \Big )c_{i}^{2} \nonumber \\
&\quad+ \Big( -\frac{805 \mu_{4}}{198}-\frac{245 \mu_{6}}{792}+\frac{8 \epsilon_{5}}{9} \Big) c_{i}^{4} + \frac{5 \mu_{6}}{24} c_{i}^{6} \bigg) \nonumber \\
&\quad + \nu^{2} \bigg(  -\frac{2047 \mu_{2}}{1512}+\frac{19 \mu_{4}}{462}+\frac{17 \mu_{6}}{2376}-\frac{73 \epsilon_{3}}{27}-\frac{2 \epsilon_{5}}{27}  \nonumber \\
&\quad+ \Big( -\frac{2047 \mu_{2}}{1512}-\frac{19 \mu_{4}}{77}-\frac{289 \mu_{6}}{2376}+\frac{146 \epsilon_{3}}{27}+\frac{22 \epsilon_{5}}{27}\Big) c_{i}^{2} \nonumber \\
&\quad+ \Big( \frac{19 \mu_{4}}{66}+\frac{245 \mu_{6}}{792}-\frac{8 \epsilon_{5}}{9} \Big)  c_{i}^{4} - \frac{5 \mu_{6}}{24} c_{i}^{6} \bigg) \Bigg] \nonumber \\
&\quad+ \pi  s_{i} \, \delta \cos 3 \Psi \Bigg[ \frac{27}{8} \left(1+c_{i}^{2}\right) \mu_{3} \Bigg] \nonumber \\ 
&\quad+ \cos 4 \Psi \Bigg[  \frac{1186 \mu_{4}}{165}-\frac{16 \mu_{6}}{495}+\frac{32 \epsilon_{5}}{45} + \Big( \frac{16 \mu_{6}}{45}-\frac{32 \epsilon_{5}}{9} \Big) c_{i}^{2} \nonumber \\ 
&\quad+ \Big( -\frac{1186 \mu_{4}}{165}-\frac{688 \mu_{6}}{495}+\frac{128 \epsilon_{5}}{45} \Big) c_{i}^{4} + \frac{16 \mu_{6}}{15} c_{i}^{6}  \nonumber \\
&\quad+ \nu \bigg( -\frac{2546 \mu_{4}}{99}+\frac{16 \mu_{6}}{99}-\frac{32 \epsilon_{5}}{9} + \Big( \frac{160 \epsilon_{5}}{9}-\frac{16 \mu_{6}}{9} \Big) c_{i}^{2}  \nonumber \\
&\quad+ \Big( \frac{2546 \mu_{4}}{99}+\frac{688 \mu_{6}}{99}-\frac{128 \epsilon_{5}}{9} \Big) c_{i}^{4} - \frac{16 \mu_{6}}{3} c_{i}^{6} \bigg) \nonumber \\
&\quad+ \nu^{2} \bigg( \frac{350 \mu_{4}}{33}-\frac{16 \mu_{6}}{99}+\frac{32 \epsilon_{5}}{9} + \Big( \frac{16 \mu_{6}}{9}-\frac{160 \epsilon_{5}}{9}\Big) c_{i}^{2} \nonumber \\
&\quad+ \Big( -\frac{350 \mu_{4}}{33}-\frac{688 \mu _{6}}{99}+\frac{128 \epsilon_{5}}{9} \Big) c_{i}^{4} + \frac{16 \mu_{6}}{3} c_{i}^{6} \bigg) \Bigg] \nonumber \\
&\quad+ \cos 6 \Psi \Bigg[ -\frac{81}{40} s_{i}^{4} \left(1+c_{i}^{2}\right)  \big(1 - 5 \nu + 5 \nu^{2}\big) \mu_{6}  \Bigg] \nonumber \\
&\quad+ s_{i} \, \delta \sin \Psi \Bigg[  -\frac{7 \mu_{3}}{120}-\frac{\mu_{3}}{12}  {\rm ln} \,2+\frac{\epsilon_{2}}{3}+\frac{4 \epsilon_{2}}{3} {\rm ln}\, 2 + \bigg( \frac{7 \mu_{3}}{40}+\frac{\mu_{3}}{4}  {\rm ln} \,2 \bigg) c_{i}^{2} \Bigg] \nonumber \\
&\quad+ s_{i} \, \delta \sin 3 \Psi \Bigg[ \bigg(- \frac{189 \mu_{3}}{40} + \frac{27 \mu_{3}}{4} {\rm ln} (3/2) \bigg) \left(1+c_{i}^{2}\right) \Bigg] \, ,\\
\xi_{+}^{(2,{\rm SO})} &= s_{i} \cos \Psi \Bigg[ \bigg( -\frac{\mu_{3}}{12} + \frac{\epsilon_{2}}{3} +  \frac{\mu_{3}}{4}  c_{i}^{2} \bigg) \Big(\boldsymbol{\chi}_a\cdot\mathbf{\hat{L}}_{\rm N} + \delta \, \boldsymbol{\chi}_s\cdot\mathbf{\hat{L}}_{\rm N}\Big) \nonumber \\
&\quad+ \nu \bigg( \Big( \frac{11 \mu_{3}}{48} - \frac{205 \epsilon_{2}}{21} - \frac{\epsilon_{4}}{112} + \big(- \frac{11 \mu_{3}}{16} + \frac{\epsilon_{4}}{16} \big) c_{i}^{2} \Big) \boldsymbol{\chi}_a\cdot\mathbf{\hat{L}}_{\rm N} \nonumber \\
&\quad+ \Big(  \frac{13 \mu_{3}}{48} - \frac{11 \epsilon_{2}}{7} + \frac{\epsilon_{4}}{112} - \big(\frac{13 \mu_{3}}{16} +  \frac{\epsilon_{4}}{16} \big) c_{i}^{2} \Big) \delta \, \boldsymbol{\chi}_s\cdot\mathbf{\hat{L}}_{\rm N} \bigg) \Bigg] \nonumber \\
&\quad+ s_{i} \cos 3 \Psi \Bigg[ - \frac{9}{4} \big(1+c_{i}^{2}\big) \Big(\boldsymbol{\chi}_a\cdot\mathbf{\hat{L}}_{\rm N} + \delta \, \boldsymbol{\chi}_s\cdot\mathbf{\hat{L}}_{\rm N}\Big) \mu_{3} \nonumber \\ 
&\quad + \nu \bigg( \Big( \frac{171 \mu_{3}}{16} + \frac{27 \epsilon_{4}}{16} + \big(\frac{171 \mu_{3}}{16} - \frac{81 \epsilon_{4}}{16} \big) c_{i}^{2} \Big) \boldsymbol{\chi}_a\cdot\mathbf{\hat{L}}_{\rm N} \nonumber \\ 
&\quad+ \Big(  \frac{45 \mu_{3}}{16} - \frac{27 \epsilon_{4}}{16} + \big(\frac{45 \mu_{3}}{16} +  \frac{81 \epsilon_{4}}{16} \big) c_{i}^{2} \Big) \delta \, \boldsymbol{\chi}_s\cdot\mathbf{\hat{L}}_{\rm N} \bigg) \Bigg] \, ,\\
\xi_{+}^{(2,{\rm SS})} &= \cos 2 \Psi \Bigg[- \left(1+c_{i}^{2}\right)  \Big( \boldsymbol{\chi}_s 
+ \delta\,\boldsymbol{\chi}_a \Big)^{2} \mu_{2}\Bigg] \, ,
\end{align}
\end{subequations}

\begin{subequations}
\label{eq:HcTD}
\begin{align}
\xi_{\times}^{(0,{\rm NS})} &= -2 c_{i} \, \mu_2 \sin 2 \Psi \, ,\\
\xi_{\times}^{(1/2,{\rm NS})} &= s_{i} c_{i} \, \delta \Bigg[ -\bigg( \frac{2 \epsilon_{2}}{3} + \frac{\mu_{3}}{12}\bigg) \sin \Psi + \frac{9 \mu_{3}}{4} \sin 3 \Psi \Bigg] \, , \\
\xi_{\times}^{(1,{\rm NS})} &= c_{i} \sin 2 \Psi \Bigg[  \frac{107 \mu_{2}}{21}+\frac{5 \mu _{4}}{21}+\frac{\epsilon_{3}}{3}  + \Big( -\frac{\mu_{4}}{3}-\epsilon_{3} \Big) c_{i}^{2} \nonumber \\
&\quad+ \nu \bigg( - \frac{55 \mu_{2}}{21}-\frac{5 \mu_{4}}{7}-\epsilon_{3} + \Big( \mu_{4}+3 \epsilon_{3} \Big) c_{i}^{2} \bigg) \Bigg] \nonumber \\
&\quad+ c_{i} s_{i}^{2} \sin 4 \Psi \Bigg[ -\frac{8}{3} \Big(1 - 3 \nu \Big) \mu_{4} \Bigg] \, ,\\
\xi_{\times}^{(1,{\rm SO})} &= s_{i} c_{i} \sin \Psi  \Bigg[    \left(\boldsymbol{\chi}_a\cdot\mathbf{\hat{L}}_{\rm N} + \delta \, \boldsymbol{\chi}_s\cdot\mathbf{\hat{L}}_{\rm N}\right) \epsilon_{2}  \Bigg] \, ,\\
\xi_{\times}^{(3/2,{\rm NS})} &= s_{i} c_{i} \, \delta \sin \Psi \Bigg[ \frac{2 \mu_{3}}{9}+\frac{\mu_{5}}{1440}+\frac{17 \epsilon_{2}}{42}+\frac{\epsilon_{4}}{35} - \Big( \frac{\mu_{5}}{480}+\frac{\epsilon_{4}}{20} \Big) c_{i}^{2} \nonumber \\
&\quad+ \nu \bigg( \frac{\mu_{3}}{18}-\frac{\mu_{5}}{720}-\frac{10 \epsilon_{2}}{21}-\frac{2 \epsilon_{4}}{35} + \Big( \frac{\mu_{5}}{240}+\frac{\epsilon_{4}}{10}\Big) c_{i}^{2} \bigg)  \Bigg] \nonumber \\
&\quad- 4 \pi c_{i} \mu_{2} \sin 2 \Psi \nonumber \\
&\quad + s_{i} c_{i} \, \delta \sin 3 \Psi \Bigg[ - 9 \mu_{3} - \frac{27 \mu_{5}}{64} + \Big( \frac{243 \mu_{5}}{320} + \frac{27 \epsilon_{4}}{20} \Big) c_{i}^{2} \nonumber \\
&\quad + \nu \bigg( \frac{9 \mu_{3}}{2}+\frac{27 \mu_{5}}{32} - \Big( \frac{243 \mu_{5}}{160}+\frac{27 \epsilon_{4}}{10}\Big) c_{i}^{2} \bigg) \Bigg] \nonumber \\
&\quad + s_{i} c_{i} \, \delta \sin 5 \Psi \Bigg[ \frac{625}{192} \big( 1 - 2\nu \big) \mu_{5} s_{i}^{2} \Bigg] \, , \\
\xi_{\times}^{(3/2,{\rm SO})} &= c_{i} \sin 2 \Psi \Bigg[\frac{4}{3} \bigg( 2 \left(\boldsymbol{\chi}_s\cdot\mathbf{\hat{L}}_{\rm N} 
+ \delta\,\boldsymbol{\chi}_a\cdot\mathbf{\hat{L}}_{\rm N}\right) \mu_{2} - \nu \Big(  2 \mu_{2} - \epsilon_{3} + 3 \epsilon_{3} c_{i}^{2} \Big) \boldsymbol{\chi}_s\cdot\mathbf{\hat{L}}_{\rm N}  \bigg) \Bigg] \, ,\\
\xi_{\times}^{(2,{\rm NS})} &= s_{i} c_{i} \, \delta \cos \Psi \Bigg[ -\frac{7 \mu_{3}}{60}-\frac{\epsilon_{2}}{3} - \bigg(\frac{\mu_{3}}{6} + \frac{4 \epsilon_{2}}{3} \bigg) {\rm ln} \, 2  \Bigg] \nonumber \\
&\quad+ s_{i} c_{i} \, \delta \cos 3 \Psi \Bigg[ \bigg(\frac{189}{20} - \frac{27}{2} {\rm ln} (3/2) \bigg) \mu_{3} \Bigg] \nonumber \\
&\quad- s_{i} c_{i} \, \delta \sin \Psi \Bigg[ \bigg(\frac{\mu_{3}}{12} + \frac{2 \epsilon_{2}}{3} \bigg) \pi  \Bigg] \nonumber \\
&\quad + c_{i} \sin 2 \Psi \Bigg[ \frac{2173 \mu_{2}}{756}-\frac{437 \mu_{4}}{462}-\frac{37 \mu_{6}}{5940}-\frac{193 \epsilon_{3}}{270}-\frac{2 \epsilon_{5}}{27} \nonumber \\
&\quad + \Big( \frac{437 \mu_{4}}{330}+\frac{31 \mu_{6}}{990}+\frac{193 \epsilon_{3}}{90}+\frac{4 \epsilon_{5}}{15} \Big) c_{i}^{2} - \Big(\frac{\mu_{6}}{36} + \frac{2 \epsilon_{5}}{9} \Big) c_{i}^{4} \nonumber \\
&\quad +  \nu \bigg( \frac{1069 \mu_{2}}{108}+\frac{575 \mu_{4}}{198}+\frac{37 \mu_{6}}{1188}+\frac{145 \epsilon_{3}}{54}+\frac{10 \epsilon_{5}}{27} \nonumber \\
&\quad - \Big( \frac{805 \mu_{4}}{198}+\frac{31 \mu_{6}}{198}+\frac{145 \epsilon_{3}}{18}+\frac{4 \epsilon_{5}}{3} \Big) c_{i}^{2} + \Big( \frac{5 \mu_{6}}{36}+\frac{10 \epsilon_{5}}{9}\Big) c_{i}^{4} \bigg) \nonumber \\
&\quad + \nu^{2} \bigg( - \frac{2047 \mu_{2}}{756}-\frac{95 \mu_{4}}{462}-\frac{37 \mu_{6}}{1188}-\frac{73 \epsilon_{3}}{54}-\frac{10 \epsilon_{5}}{27} \nonumber \\
&\quad + \Big( \frac{19 \mu_{4}}{66}+\frac{31 \mu_{6}}{198}+\frac{73 \epsilon_{3}}{18}+\frac{4 \epsilon_{5}}{3} \Big) c_{i}^{2} - \Big( \frac{10 \epsilon_{5}}{9}+\frac{5 \mu_{6}}{36} \Big) c_{i}^{4} \bigg)
\Bigg] \nonumber \\
&\quad + s_{i} c_{i} \, \delta \sin 3 \Psi \Bigg[ \frac{27 \pi }{4} \mu_{3} \Bigg] \nonumber \\
&\quad + c_{i}  \sin 4  \Psi \Bigg[ \frac{2372 \mu_{4}}{165}+\frac{64 \mu_{6}}{99}-\frac{16 \epsilon_{5}}{45} - \Big( \frac{2372 \mu_{4}}{165}+\frac{1024 \mu_{6}}{495}+\frac{64 \epsilon_{5}}{45} \Big) c_{i}^{2}\nonumber \\
&\quad + \Big( \frac{64 \mu_{6}}{45}+\frac{16 \epsilon_{5}}{9} \Big) c_{i}^{4} \nonumber \\
&\quad + \nu \bigg( -\frac{5092 \mu_{4}}{99}-\frac{320 \mu_{6}}{99}+\frac{16 \epsilon_{5}}{9} + \Big( \frac{5092 \mu_{4}}{99}+\frac{1024 \mu_{6}}{99}+\frac{64 \epsilon_{5}}{9} \Big) c_{i}^{2} \nonumber \\
&\quad- \Big( \frac{80 \epsilon_{5}}{9} + \frac{64 \mu_{6}}{9} \Big) c_{i}^{4} \bigg) \nonumber \\
&\quad+ \nu^{2} \bigg( \frac{700 \mu_{4}}{33}+\frac{320 \mu_{6}}{99}-\frac{16 \epsilon_{5}}{9}  - \Big( \frac{700 \mu_{4}}{33}+\frac{1024 \mu_{6}}{99}+\frac{64 \epsilon_{5}}{9} \Big) c_{i}^{2} \nonumber \\
&\quad+ \Big( \frac{64 \mu_{6}}{9}+\frac{80 \epsilon_{5}}{9} \Big) c_{i}^{4} \bigg) \Bigg] \nonumber \\
&\quad+ c_{i}  \sin 6  \Psi \Bigg[ - \frac{81}{20} s_{i}^{4} \bigg(1 - 5 \nu + 5 \nu ^2 \bigg) \mu_{6} \Bigg] \, , \\
\xi_{\times}^{(2,{\rm SO})} &= s_{i} c_{i} \sin \Psi \Bigg[ \Big( \frac{\mu_{3}}{6}+\frac{\epsilon_{2}}{3} \Big) \left(\boldsymbol{\chi}_a\cdot\mathbf{\hat{L}}_{\rm N} 
+ \delta\,\boldsymbol{\chi}_s\cdot\mathbf{\hat{L}}_{\rm N}\right) \nonumber \\ 
&\quad + \nu \bigg( \Big(-\frac{11 \mu_{3}}{24}-\frac{205 \epsilon_{2}}{21}-\frac{\epsilon_{4}}{14} + \frac{\epsilon_{4}}{8} c_{i}^{2} \Big) \boldsymbol{\chi}_a\cdot\mathbf{\hat{L}}_{\rm N} 
\nonumber \\ 
&\quad - \Big( \frac{13 \mu_{3}}{24}+\frac{11 \epsilon_{2}}{7}-\frac{\epsilon_{4}}{14}  + \frac{\epsilon_{4}}{8} c_{i}^{2} \Big) \delta \, \boldsymbol{\chi}_s\cdot\mathbf{\hat{L}}_{\rm N} \bigg) \Bigg] \nonumber \\
&\quad + s_{i} c_{i} \sin 3 \Psi \Bigg[ - \frac{9 \mu_{3}}{2} \left(\boldsymbol{\chi}_a\cdot\mathbf{\hat{L}}_{\rm N} 
+ \delta\,\boldsymbol{\chi}_s\cdot\mathbf{\hat{L}}_{\rm N}\right)  \nonumber \\
&\quad+ \nu \bigg( \Big( \frac{171 \mu_{3}}{8} - \frac{27 \epsilon_{4}}{8} c_{i}^{2} \Big) \boldsymbol{\chi}_a\cdot\mathbf{\hat{L}}_{\rm N} + \Big( \frac{45 \mu_{3}}{8} + \frac{27 \epsilon_{4}}{8} c_{i}^{2} \Big) \delta\,\boldsymbol{\chi}_s\cdot\mathbf{\hat{L}}_{\rm N} \bigg)
\Bigg] \, , \\
\xi_{\times}^{(2,{\rm SS})} &= \sin 2 \Psi \Bigg[ - 2 c_{i}  \Big( \boldsymbol{\chi}_s + \delta\,\boldsymbol{\chi}_a \Big)^{2} \mu_{2} \Bigg] \, .
\end{align}
\end{subequations}
The symbols $c_i$ and $s_{i}$ are the shorthand for the cosine and sine of the inclination angle, respectively, i.e.,  $c_i=\cos i$, $s_{i}=\sin i$. The variable $\Psi$ is the shifted orbital phase $\Psi=\Phi - 2 \, (v^{3}/c^{3}) \, {\rm ln} (\omega/ \omega_{0}) $, where $\omega_{0}$ is an arbitrary reference orbital frequency. In the MPM-PN formalism~\cite{BDI95}, the arbitrary frequency $\omega_{0}$ parametrizes a certain freedom to construct the (Bondi-type) radiative coordinates in the far zone. More precisely, it is linked to the difference of origins of the time coordinates in the far zone (radiative coordinates) and in the near zone (harmonic coordinates) [see Eq.~(5) and the following section of Ref.~\cite{BIWW96} and Eq.~(4.36) and the following section of Ref.~\cite{ABIQ04} for a comprehensive explanation]. It should be noted here that we do not include the nonlinear memory effect,\footnote{The nonlinear memory effect is a nonoscillatory low-frequency contribution to the gravitational wave amplitude with poor observational prospects for the ground-based interferometers. This leads to a negligible impact on the measurement of different multipole parameters and thus ignored.} which first appears in the polarization at the Newtonian order \cite{ABIQ04}. As an algebraic check, in the limit $\mu_l\rightarrow 1, \epsilon_l\rightarrow 1$, the nonspinning contributions to the parametrized polarizations reduce to Eq.~(3) and Eq.~(4) of Ref.~\cite{BIWW96}. In the same limit $\mu_l\rightarrow 1, \epsilon_l\rightarrow 1$, we also confirm the recovery of the GR results for spin-orbit and spin-spin contributions to the polarizations reported in Ref.~\cite{MKAF2016}. A complete list of time-domain amplitude coefficients contributing at the 2PN order is provided in a separate file (\textbf{\emph{supl-mk23.m}}) for convenient use, and is readable in \textit{Mathematica}.

As is evident from the structure, the even-parity multipole moments [i.e., mass-type moments with even $l$ (e.g., $\mu_{2}$, $\mu_{4}$, etc.) and current-type moments with odd $l$ (e.g., $\epsilon_{3}$, $\epsilon_{5}$, etc.)] contribute to terms associated with even harmonics, i.e., $2\Psi$, $4\Psi$, etc., whereas, odd-parity multipole moments [i.e., mass-type moments with odd $l$ (e.g., $\mu_{3}$, $\mu_{5}$, etc.) and current-type moments with even $l$ (e.g., $\epsilon_{2}$, $\epsilon_{4}$, etc.)] contribute to terms associated with odd harmonics, i.e., $\Psi$, $3\Psi$, etc. Moreover, the deviation parameters associated to the mass-type moments, $\mu_l$, contribute to the harmonics $l\Psi$, $(l-2)\Psi, \cdots$, and the parameters attached to the current-type moments, $\epsilon_l$ contribute harmonics $(l-1)\Psi$, $(l-3)\Psi, \cdots$.  The tail contributions appear in the polarization at 1.5PN, and 2PN orders, are due to the 1.5PN tail integral performed on mass-quadrupole (and so $\mu_2$ appears at 1.5PN), mass-octupole ($\mu_3$ appearing at 2PN), and current-quadrupole ($\epsilon_2$ at 2PN). Spin-orbit corrections to the amplitude first arise at 1PN order due to spin-dependent terms in the current quadrupole moment ($\epsilon_2$) at 0.5PN order. The 2PN quadratic-in-spin corrections in the amplitude come from 2PN quadratic-in-spin terms in the mass-type quadrupole moment.

\section {Frequency Domain waveform}\label{sec:FD-waveform}

The majority of gravitational wave data analysis
is performed in the frequency domain.
Hence, significant effort has gone into developing frequency-domain waveform models. 
Therefore, using the time-domain polarization computed in the previous section, we also provide the corresponding frequency domain parametrized multipolar polarizations.

We follow Ref.~\cite{ABFO08} to compute the same. The basic scheme for the computation of the frequency domain polarizations involves two steps. In the first step, one needs to collect terms in the time domain polarizations by PN order and by sines or cosines of harmonics of the orbital frequency and write it in a compact form. In the next step, to compute the Fourier transform of the time domain polarizations, one needs to apply the stationary-phase approximation (SPA) \cite{CF94, DIS00} to each term associated with different harmonics of the orbital phase. The SPA is generally used to compute the Fourier transform of a time domain signal in the form $A(t)e^{i\phi(t)}$ with the amplitude [$A(t)$] varying slowly with time (i.e., $d \, {\rm ln} A(t)/dt \ll d \phi(t)/dt$) whereas, the phase is rapidly oscillating [i.e., $\frac{d^{2}\phi(t)}{dt^{2}}\ll (\frac{d\phi(t)}{dt})^{2}$]~\cite{Sintes:1999cg,VanDenBroeck:2006qi,VanDenBroeck:2006qu,VanDenBroeck:2006ar}. Since the quantities $(d \, {\rm ln} A(t)/dt)/(d \phi(t)/dt)$ and $(d^{2}\phi(t)/dt^{2})/((d\phi(t)/dt)^{2})$ are very small compared to unity for the entire inspiral of a compact binary coalescence in GR provides the justification for using the SPA while computing the frequency-domain waveform models. We refer to Sec.~VI B of Ref.~\cite{ABFO08} for more details. Since the time-domain parametrized multipolar waveform model also respects the criteria of SPA, we follow the same while computing the frequency domain polarizations at 2PN, which reads,
\begin{align}
    \Tilde{h}_{+,\times} (f) = \frac{G^{2}M^{2}}{c^{5}D_{L}}\sqrt{\frac{5 \, \pi \, \nu}{48}} \sum_{n=0}^{4} \sum_{k=1}^{6} V_{k}^{n-7/2} \, H_{+,\times}^{(k,n)} \,  e^{i \big( k \Psi_{\rm SPA}(f/k)-\pi/4\big)} \, .
\end{align}

Here, $\Tilde{h}_{+,\times} (f)$ denote the parametrized multipolar polarizations in the frequency domain as observed by the detector, while $f$ denotes the GW frequency, and 
$k$ indicates the harmonics of the orbital phase. For the $k$th harmonic, the dimensionless gauge invariant PN parameter $v/c\equiv v(t)/c$, entering in the time domain polarizations, is replaced by $V_{k}$, defined as $V_{k}\equiv (2 \, \pi \, G \, M \, f/c^{3}\, k)^{1/3}$. The function $\Psi_{\rm SPA}(f)$ represents the parametrized multipolar GW phasing for the first harmonic in the frequency domain as obtained under the SPA (see Secs. VI B and VI C of Ref.~\cite{ABFO08} for a detailed discussion). The 3.5PN parametrized multipolar GW phase of the second harmonic for nonspinning compact binaries can be found in Eq.~(2.16) of Ref.~\cite{Kastha:2018bcr}, whereas the phasing of the second harmonic for the nonprecessing spinning compact binaries is given in Eq.~(3.9) and Eq.~(3.10) of Ref.~\cite{Kastha:2019}. Lastly, the coefficients $H_{+,\times}^{(k,n)}$ depend on the masses, the spins, and the orbital inclination angle of the compact binary and contains the multipole parameters.\footnote{The amplitude coefficients $\mathcal{C}_{k}^{n}$ in Refs.~\cite{ABFO08,MKAF2016} [see Eq.~(6.13) in Ref.~\cite{ABFO08} and Eq.~(1) in Ref.~\cite{MKAF2016}] are related to $H_{+,\times}^{(k,n)}$ via the relation $\mathcal{C}_{k}^{n}=H_{+}^{(k,n)} F_+ + H_{\times}^{(k,n)} F_\times $, where $F_{+,\times}$ are the antenna pattern functions of the GW detector to the two different polarizations.}

We provide the explicit expressions of different nonvanishing $H_{+,\times}^{(k,n)}$ in terms of scaled multipolar coefficients $\hat{\mu}_{l}(\equiv \mu_{l}/\mu_{2})$ and $\hat{\epsilon}_{l}(\equiv \epsilon_{l}/\mu_{2})$, and they read as
\begin{subequations}
\label{eq:HpFD}
\begin{align}
H_{+}^{(2,0)} &= - \frac{1}{\sqrt{2}} \Big[ \left(1+c_{i}^2\right) \Big] \, \Theta (2 \, F_{\rm cut}-f)\, , \\
H_{+}^{(1,1)} &= s_{i} \, \delta \bigg[ \frac{\hat{\mu}_{3}}{24}-\frac{2 \hat{\epsilon}_{2}}{3} -\frac{\hat{\mu}_{3}}{8} c_{i}^2 \bigg] \, \Theta (F_{\rm cut}-f)\, , \\
H_{+}^{(3,1)} &= \frac{1}{\sqrt
{3}} \bigg[ \frac{9}{8} \, s_{i} \, \delta \, (1+c_{i}^2) \, \hat{\mu}_{3} \bigg] \, \Theta (3 \, F_{\rm cut}-f)\, , \\
H_{+}^{(1,2)} &= s_{i}  \,  \bigg[\left(\boldsymbol{\chi}_a\cdot\mathbf{\hat{L}}_{\rm N} + \delta \, \boldsymbol{\chi}_s\cdot\mathbf{\hat{L}}_{\rm N}\right) \hat{\epsilon}_{2} \bigg] \, \Theta (F_{\rm cut}-f)\, , \\
H_{+}^{(2,2)} &= \frac{1}{\sqrt{2}} \Bigg[  \frac{3}{4} + \frac{1367 \hat{\mu}_{3}^2}{2016} -\frac{\hat{\mu}_{4}}{21}+\frac{\hat{\epsilon}_{2}^2}{72}+\frac{2 \hat{\epsilon}_{3}}{3} + \left(\frac{1}{12} - \frac{1367 \hat{\mu}_{3}^2}{504} + \frac{\hat{\mu}_{4}}{7}-\frac{\hat{\epsilon}_{2}^2}{18}-2 \hat{\epsilon}_{3} \right) \nu \nonumber \\
&\quad + \Bigg( \frac{3}{4} + \frac{1367 \hat{\mu}_{3}^2}{2016}+\frac{2 \hat{\mu}_{4}}{7}+\frac{\hat{\epsilon}_{2}^2}{72}-\frac{4 \hat{\epsilon}_{3}}{3} + \left( \frac{1}{12} -\frac{1367 \hat{\mu}_{3}^2}{504}-\frac{6 \hat{\mu}_{4}}{7}-\frac{\hat{\epsilon}_{2}^2}{18}+4 \hat{\epsilon}_{3} \right) \nu \Bigg) c_{i}^{2} \nonumber \\
&\quad + \left(\nu -\frac{1}{3}\right) \hat{\mu}_{4} \, c_{i}^{4}
\Bigg] \Theta (2 \, F_{\rm cut}-f)\, , \\
H_{+}^{(4,2)} &= - \frac{1}{2} \, s_{i}^{2} \,  \bigg[ \frac{4}{3} (1-3\nu) \, (1 + c_{i}^{2}) \, \hat{\mu}_{4} \bigg] \, \Theta (4 \, F_{\rm cut}-f)\, , \\
H_{+}^{(1,3)} &= s_{i} \, \delta \, \bigg[ -\frac{73 \hat{\mu} _{3}}{2016}-\frac{\hat{\mu}_{3} \hat{\epsilon}_{2}^2}{1728}+\frac{1367 \hat{\mu}_{3}^2 \hat{\epsilon}_{2}}{3024}-\frac{1367 \hat{\mu}_{3}^3}{48384}-\frac{\hat{\mu}_{5}}{2880}-\frac{50 \hat{\epsilon}_{2}}{63}+\frac{\hat{\epsilon}_{2}^3}{108}+\frac{\hat{\epsilon}_{4}}{280}  \nonumber \\
&\quad + \left( -\frac{173 \hat{\mu}_{3}}{2016} + \frac{\hat{\mu}_{3} \hat{\epsilon}_{2}^2}{432} -\frac{1367 \hat{\mu}_{3}^2 \hat{\epsilon}_{2}}{756} + \frac{1367 \hat{\mu}_{3}^3}{12096}+\frac{\hat{\mu}_{5}}{1440} +\frac{19 \hat{\epsilon}_{2}}{42}-\frac{\hat{\epsilon}_{2}^3}{27}-\frac{\hat{\epsilon}_{4}}{140} \right) \nu \nonumber \\
&\quad + \Bigg( \frac{73 \hat{\mu}_{3}}{672} +\frac{\hat{\mu}_{3} \hat{\epsilon}_{2}^2}{576} + \frac{1367 \hat{\mu}_{3}^3}{16128}+\frac{\hat{\mu}_{5}}{240}-\frac{\hat{\epsilon}_{4}}{40}  + \left(\frac{173 \hat{\mu}_{3}}{672} -\frac{\hat{\mu}_{3} \hat{\epsilon}_{2}^2}{144} -\frac{1367 \hat{\mu}_{3}^3}{4032}-\frac{\hat{\mu}_{5}}{120}+\frac{\hat{\epsilon}_{4}}{20} \right)\nu \Bigg) c_{i}^{2} \nonumber \\
&\quad+ \left( -\frac{\hat{\mu}_{5}}{192} + \frac{\hat{\mu}_{5}}{96} \nu \right)c_i^{4}\bigg] \, \Theta (F_{\rm cut}-f)\, , \\
H_{+}^{(2,3)} &= \frac{1}{\sqrt{2}} \Bigg[ -\left(1+c_{i}^2\right)  \left( \frac{10}{3} + \frac{\hat{\epsilon}_{2}^{2}}{24} \right) \delta \left(\boldsymbol{\chi}_a\cdot\mathbf{\hat{L}}_{\rm N} \right) + \Bigg( -\left(1+c_{i}^2\right)  \left( \frac{10}{3} + \frac{\hat{\epsilon}_{2}^{2}}{24} \right)\, \nonumber \\
&\quad + \bigg( \frac{1}{6} \left(1+c_{i}^2\right) \hat{\epsilon}_{2}^{2} +  \frac{5}{3} \left(1+c_{i}^2\right) + \frac{8}{3} \left(1-2 c_{i}^2\right) \hat{\epsilon}_{3}  \bigg) \nu \Bigg)  \left(\boldsymbol{\chi}_s\cdot\mathbf{\hat{L}}_{\rm N} \right) \Bigg]\, \Theta (2 \, F_{\rm cut}-f)\, , \\
H_{+}^{(3,3)} &= \frac{1}{\sqrt{3}} s_{i} \, \delta \Bigg[ -\frac{555 \hat{\mu}_{3}}{224} -\frac{\hat{\mu}_{3} \hat{\epsilon}_{2}^2}{64} -\frac{1367 \hat{\mu}_{3}^3}{1792}+\frac{27 \hat{\mu}_{5}}{640}-\frac{27 \hat{\epsilon}_{4}}{40}  + \left(\frac{153 \hat{\mu}_{3}}{224} + \frac{\hat{\mu}_{3} \hat{\epsilon}_{2}^2}{16} + \frac{1367 \hat{\mu}_{3}^3}{448}-\frac{27 \hat{\mu}_{5}}{320}+\frac{27 \hat{\epsilon}_{4}}{20}\right)\nu \, \nonumber \\ 
&\quad+ \Bigg(  -\frac{555 \hat{\mu}_{3}}{224} - \frac{\hat{\mu}_{3} \hat{\epsilon}_{2}^2}{64 } -\frac{1367 \hat{\mu}_{3}^3}{1792}-\frac{27 \hat{\mu}_{5}}{80} +\frac{81 \hat{\epsilon}_{4}}{40} + \left( \frac{153 \hat{\mu}_{3}}{224} + \frac{\hat{\mu}_{3} \hat{\epsilon}_{2}^2}{16}+ \frac{1367 \hat{\mu}_{3}^3}{448}+\frac{27 \hat{\mu}_{5}}{40}-\frac{81 \hat{\epsilon}_{4}}{20} \right) \nu \Bigg) c_{i}^{2} \, \nonumber \\
&\quad+ \frac{81}{128} \left( 1 - 2\nu \right) \hat{\mu}_5 c_{i}^{4} \Bigg] \, \Theta (3 \, F_{\rm cut}-f)\, , \\
H_{+}^{(5,3)} &= \frac{1}{\sqrt{5}} s_{i}^{3} \, \delta \, (1-2\nu) \Bigg[ \frac{625}{384} (1+c_{i}^{2}) \hat{\mu}_{5} \Bigg] \, \Theta (5 \, F_{\rm cut}-f)\, ,\\
H_{+}^{(1,4)} &= s_{i} \Bigg[ \delta \Bigg( i \Big(-\frac{7 \hat{\mu}_{3}}{120}+\frac{\hat{\epsilon}_{2}}{3}\Big) + \Big( - \frac{\hat{\mu}_{3}}{24} + \frac{2 \hat{\epsilon}_{2}}{3} \Big) \pi + i \Big(- \frac{\hat{\mu}_{3}}{12}  + \frac{4 \hat{\epsilon}_{2}}{3}\Big) {\rm ln} \, 2   + \Big(\frac{\pi}{8}+\frac{7 i}{40}+\frac{i}{4} {\rm ln} \, 2\Big) \hat{\mu}_{3} \, c_{i}^{2} \Bigg) \, \nonumber \\
&\quad+  \Bigg( \frac{\hat{\mu}_{3}}{9}+\frac{\hat{\mu}_{3} \hat{\epsilon}_{2}^{2}}{576}-\frac{1367 \hat{\mu}_{3}^{2} \hat{\epsilon}_{2}}{2016}-\frac{247 \hat{\epsilon}_{2}}{252} -\frac{\hat{\epsilon}_{2}^{3}}{24} + \left( \frac{7 \hat{\mu}_{3}}{48}-\frac{\hat{\mu}_{3} \hat{\epsilon}_{2}^{2}}{144}+\frac{1367 \hat{\mu}_{3}^{2} \hat{\epsilon}_{2}}{504}-\frac{27 \hat{\epsilon}_{2}}{28}+\frac{\hat{\epsilon}_{2}^{3}}{6}+\frac{\hat{\epsilon}_{4}}{112} \right) \nu  \, \nonumber \\
&\quad+ \bigg( - \frac{\hat{\mu}_{3}}{3} - \frac{\hat{\mu}_{3}\hat{\epsilon}_{2}^{2}}{192}  + \Big( -\frac{7 \hat{\mu}_{3}}{16} + \frac{\hat{\mu}_{3} \hat{\epsilon}_{2}^{2}}{48} -\frac{\hat{\epsilon_{4}}}{16} \Big) \nu  \bigg) c_{i}^{2} \Bigg) \delta \left(\boldsymbol{\chi}_s\cdot\mathbf{\hat{L}}_{\rm N} \right) \, \nonumber \\
&\quad+  \Bigg( \frac{\hat{\mu}_{3}}{9}+\frac{\hat{\mu}_{3} \hat{\epsilon}_{2}^{2}}{576}-\frac{1367 \hat{\mu}_{3}^{2} \hat{\epsilon}_{2}}{2016}-\frac{\hat{\epsilon}_{2}^{3}}{24}-\frac{247 \hat{\epsilon}_{2}}{252}  + \left( -\frac{79 \hat{\mu}_{3}}{144}-\frac{\hat{\mu}_{3} \hat{\epsilon}_{2}^{2}}{144}+\frac{1367 \hat{\mu}_{3}^{2} \hat{\epsilon}_{2}}{504}+\frac{325 \hat{\epsilon}_{2}}{252}+\frac{\hat{\epsilon}_{2}^{3}}{6}-\frac{\hat{\epsilon}_{4}}{112} \right) \nu \, \nonumber \\
&\quad+ \bigg(  -\frac{\hat{\mu}_{3}}{3}-\frac{\hat{\mu}_{3} \hat{\epsilon}_{2}^{2}}{192} + \Big( \frac{79 \hat{\mu}_{3}}{48}+\frac{\hat{\mu}_{3} \hat{\epsilon}_{2}^{2}}{48} + \frac{\hat{\epsilon}_{4}}{16} \Big) \nu  \bigg) c_{i}^{2} \Bigg) \left(\boldsymbol{\chi}_a\cdot\mathbf{\hat{L}}_{\rm N} \right) \Bigg]\, \Theta (F_{\rm cut}-f)\, , \\
H_{+}^{(2,4)} &= \frac{1}{\sqrt{2}} \Bigg[ \frac{171}{32}-\frac{419641 \hat{\mu}_{3}^{2}}{169344} + \frac{1367 \hat{\mu}_{3}^{2} \hat{\mu}_{4}}{42336}-\frac{1367 \hat{\mu}_{3}^{2} \hat{\epsilon}_{2}^{2}}{48384} -\frac{1367 \hat{\mu}_{3}^{2} \hat{\epsilon}_{3}}{3024} -\frac{1868689 \hat{\mu}_{3}^{4}}{2709504} +\frac{10049 \hat{\mu}_{4}}{97020} \, \nonumber \\ 
&\quad+ \frac{\hat{\mu}_{4} \hat{\epsilon}_{2}^2}{1512} +\frac{8965 \hat{\mu}_{4}^{2}}{7938}+\frac{17 \hat{\mu}_{6}}{11880}+\frac{263 \hat{\epsilon}_{2}^{2}}{6048}-\frac{\hat{\epsilon}_{2}^{2} \hat{\epsilon}_{3}}{108}-\frac{\hat{\epsilon}_{2}^{4}}{3456}-\frac{437 \hat{\epsilon}_{3}}{1890}+\frac{5 \hat{\epsilon}_{3}^{2}}{126}-\frac{2 \hat{\epsilon}_{5}}{135}   \, \nonumber \\
&\quad+ \bigg(-\frac{7}{2} + \frac{261073 \hat{\mu}_{3}^{2}}{24192} -\frac{1367 \hat{\mu}_{3}^{2} \hat{\mu}_{4}}{6048} + \frac{1367 \hat{\mu}_{3}^{2} \hat{\epsilon}_{2}^{2}}{6048} + \frac{1367 \hat{\mu}_{3}^{2} \hat{\epsilon}_{3}}{432} + \frac{1868689 \hat{\mu}_{3}^4}{338688}-\frac{1250 \hat{\mu}_{4}}{4851} \, \nonumber \\
&\quad-\frac{8965 \hat{\mu}_{4}^2}{1323}-\frac{\hat{\mu}_{4} \hat{\epsilon}_{2}^{2}}{216} - \frac{17 \hat{\mu}_{6}}{2376}-\frac{1159 \hat{\epsilon}_{2}^{2}}{6048}+\frac{7 \hat{\epsilon}_{2}^{2} \hat{\epsilon}_{3}}{108}+\frac{\hat{\epsilon}_{2}^{4}}{432}+\frac{160 \hat{\epsilon}_{3}}{189}-\frac{5 \hat{\epsilon}_{3}^2}{21}+\frac{2 \hat{\epsilon}_{5}}{27} \bigg) \nu \, \nonumber \\
&\quad+ \bigg(\frac{19}{288} - \frac{49649 \hat{\mu}_{3}^{2}}{14112} + \frac{1367 \hat{\mu}_{3}^{2} \hat{\mu}_{4}}{3528} - \frac{1367 \hat{\mu}_{3}^{2} \hat{\epsilon}_{2}^{2}}{3024} - \frac{1367 \hat{\mu}_{3}^{2} \hat{\epsilon}_{3}}{252} - \frac{1868689 \hat{\mu}_{3}^{4}}{169344} - \frac{1021 \hat{\mu}_{4}}{6468} \, \nonumber \\
&\quad+ \frac{\hat{\mu}_{4} \hat{\epsilon}_{2}^{2}}{126}  
+ \frac{8965 \hat{\mu}_{4}^{2}}{882} + \frac{17 \hat{\mu}_{6}}{2376} + \frac{107 \hat{\epsilon}_{2}^{2}}{1512} - \frac{\hat{\epsilon}_{2}^{2} \hat{\epsilon}_{3}}{9} - \frac{\hat{\epsilon}_{2}^{4}}{216} + \frac{31 \hat{\epsilon}_{3}}{378} + \frac{5 \hat{\epsilon}_{3}^{2}}{14} -\frac{2 \hat{\epsilon}_{5}}{27} \bigg) \nu^{2} \, \nonumber \\
&\quad+ \Bigg( \frac{171}{32} - \frac{419641 \hat{\mu}_{3}^{2}}{169344} -\frac{1367 \hat{\mu}_{3}^{2} \hat{\mu}_{4}}{7056}- \frac{1367 \hat{\mu}_{3}^{2} \hat{\epsilon}_{2}^{2}}{48384}+\frac{1367 \hat{\mu}_{3}^{2} \hat{\epsilon}_{3}}{1512}-\frac{1868689 \hat{\mu}_{3}^{4}}{2709504} - \frac{10049 \hat{\mu}_{4}}{16170} \, \nonumber \\ 
&\quad - \frac{\hat{\mu}_{4} \hat{\epsilon}_{2}^{2}}{252} 
+ \frac{8965 \hat{\mu}_{4}^{2}}{7938} -\frac{289 \hat{\mu}_{6}}{11880} + \frac{263 \hat{\epsilon}_{2}^{2}}{6048}+\frac{\hat{\epsilon}_{2}^2 \hat{\epsilon}_{3}}{54} - \frac{\hat{\epsilon}_{2}^4}{3456} +\frac{5 \hat{\epsilon}_{3}^{2}}{126}+\frac{437 \hat{\epsilon}_{3}}{945}+\frac{22 \hat{\epsilon}_{5}}{135} \, \nonumber \\
&\quad+ \bigg( - \frac{7}{2} + \frac{261073 \hat{\mu}_{3}^{2}}{24192} + \frac{1367 \hat{\mu}_{3}^{2} \hat{\mu}_{4}}{1008}+\frac{1367 \hat{\mu}_{3}^{2} \hat{\epsilon}_{2}^{2}}{6048}-\frac{1367 \hat{\mu}_{3}^{2} \hat{\epsilon}_{3}}{216}+\frac{1868689 \hat{\mu}_{3}^{4}}{338688}+ \frac{2500 \hat{\mu}_{4}}{1617}\, \nonumber \\
&\quad+\frac{\hat{\mu}_{4} \hat{\epsilon}_{2}^{2}}{36}-\frac{8965 \hat{\mu}_{4}^{2}}{1323}+\frac{289 \hat{\mu}_{6}}{2376} 
-\frac{1159 \hat{\epsilon}_{2}^{2}}{6048}-\frac{7 \hat{\epsilon}_{2}^{2} \hat{\epsilon}_{3}}{54}+\frac{\hat{\epsilon}_{2}^{4}}{432}-\frac{320 \hat{\epsilon}_{3}}{189}-\frac{5 \hat{\epsilon}_{3}^{2}}{21}-\frac{22 \hat{\epsilon}_{5}}{27} \bigg) \nu \, \nonumber \\
&\quad+ \bigg( \frac{19}{288}-\frac{49649 \hat{\mu}_{3}^{2}}{14112} -\frac{1367 \hat{\mu}_{3}^{2} \hat{\mu}_{4}}{588}-\frac{1367 \hat{\mu}_{3}^{2} \hat{\epsilon}_{2}^{2}}{3024}+\frac{1367 \hat{\mu}_{3}^{2} \hat{\epsilon}_{3}}{126}-\frac{1868689 \hat{\mu}_{3}^4}{169344}+\frac{1021 \hat{\mu}_{4}}{1078}\, \nonumber \\
&\quad -\frac{\hat{\mu}_{4} \hat{\epsilon}_{2}^{2}}{21}+\frac{8965 \hat{\mu}_{4}^{2}}{882} -\frac{289 \hat{\mu}_{6}}{2376}+\frac{107 \hat{\epsilon}_{2}^{2}}{1512}+\frac{2 \hat{\epsilon}_{2}^{2} \hat{\epsilon}_{3}}{9}-\frac{\hat{\epsilon}_{2}^{4}}{216}-\frac{31 \hat{\epsilon}_{3}}{189}+\frac{5 \hat{\epsilon}_{3}^{2}}{14}+\frac{22 \hat{\epsilon}_{5}}{27} \bigg) \nu^{2}
\Bigg) c_{i}^{2} \, \nonumber \\
&\quad+ \Bigg( \frac{1367 \hat{\mu}_{3}^{2} \hat{\mu}_{4}}{6048}+\frac{10049 \hat{\mu}_{4}}{13860}+\frac{\hat{\mu}_{4} \hat{\epsilon}_{2}^{2}}{216}+\frac{49 \hat{\mu}_{6}}{792}-\frac{8 \hat{\epsilon}_{5}}{45}  + \bigg( -\frac{1367 \hat{\mu}_{3}^{2} \hat{\mu}_{4}}{864}-\frac{1250 \hat{\mu}_{4}}{693}-\frac{7 \hat{\mu}_{4} \hat{\epsilon}_{2}^2}{216}-\frac{245 \hat{\mu}_{6}}{792}+\frac{8 \hat{\epsilon}_{5}}{9} \bigg)\nu \, \nonumber \\
&\quad+ \bigg( \frac{1367 \hat{\mu}_{3}^{2} \hat{\mu}_{4}}{504}-\frac{1021 \hat{\mu}_{4}}{924}+\frac{\hat{\mu}_{4} \hat{\epsilon}_{2}^{2}}{18}+\frac{245 \hat{\mu}_{6}}{792}-\frac{8 \hat{\epsilon}_{5}}{9} \bigg) \nu^{2} \Bigg) c_{i}^{4} - \frac{\hat{\mu}_{6}}{24} \bigg(1- 5 \nu + 5 \nu^{2} \bigg) c_{i}^{6} \, \nonumber \\ 
&\quad+ \Big(1+c_{i}^{2}\Big) \Bigg( \left(3+\frac{\hat{\epsilon}_{2}^{2}}{16} \right) \delta \left(\boldsymbol{\chi}_a\cdot\mathbf{\hat{L}}_{\rm N}\right) \left(\boldsymbol{\chi}_s\cdot\mathbf{\hat{L}}_{\rm N}\right) + \bigg( \frac{3}{2}+ \frac{\hat{\epsilon}_{2}^{2}}{32} - 6 \nu \bigg) \left(\boldsymbol{\chi}_a\cdot\mathbf{\hat{L}}_{\rm N}\right)^{2} \, \nonumber \\ 
&\quad+ \bigg( \frac{3}{2}+ \frac{\hat{\epsilon}_{2}^{2}}{32} - \frac{\hat{\epsilon}_{2}^{2}}{8} \nu \bigg) \left(\boldsymbol{\chi}_s\cdot\mathbf{\hat{L}}_{\rm N}\right)^{2} \Bigg) \Bigg]  \, \Theta (2 F_{\rm cut}-f) \, , \\
H_{+}^{(3,4)} &= \frac{1}{\sqrt{3}} s_{i} \Bigg[ \delta \left(1+c_{i}^{2}\right) \bigg( -\frac{189}{40} i + \frac{9 \pi}{8} + \frac{27}{4} i \, {\rm ln}(3/2) \bigg) \hat{\mu}_{3} +  \Bigg(  3 \hat{\mu}_{3}+\frac{3 \hat{\mu}_{3} \hat{\epsilon}_{2}^2}{64} + \left( -\frac{165 \hat{\mu}_{3}}{16}-\frac{3 \hat{\mu}_{3} \hat{\epsilon}_{2}^{2}}{16}+\frac{27 \hat{\epsilon}_{4}}{16} \right) \nu \, \nonumber \\
&\quad+ \bigg( 3 \hat{\mu}_{3}+\frac{3 \hat{\mu}_{3} \hat{\epsilon}_{2}^2}{64} + \Big( -\frac{165 \hat{\mu}_{3}}{16}-\frac{3 \hat{\mu}_{3} \hat{\epsilon}_{2}^{2}}{16}-\frac{81 \hat{\epsilon}_{4}}{16} \Big) \nu \bigg) c_{i}^{2} \Bigg) \left(\boldsymbol{\chi}_a\cdot\mathbf{\hat{L}}_{\rm N}\right) + \Bigg( 3 \hat{\mu}_{3}+\frac{3 \hat{\mu}_{3} \hat{\epsilon}_{2}^{2}}{64} + \bigg( -\frac{9 \hat{\mu}_{3}}{16}\, \nonumber \\
&\quad-\frac{3 \hat{\mu}_{3} \hat{\epsilon}_{2}^{2}}{16} - \frac{27 \hat{\epsilon}_{4}}{16} \bigg) \nu + \bigg( 3 \hat{\mu}_{3}+\frac{3 \hat{\mu}_{3} \hat{\epsilon}_{2}^{2}}{64}  + \Big( -\frac{9 \hat{\mu}_{3}}{16}-\frac{3 \hat{\mu}_{3} \hat{\epsilon}_{2}^{2}}{16}+\frac{81 \hat{\epsilon}_{4}}{16}  \Big) \nu \bigg) c_{i}^{2}  \Bigg) \delta \left(\boldsymbol{\chi}_s\cdot\mathbf{\hat{L}}_{\rm N}\right) \Bigg]  \, \Theta (3 F_{\rm cut}-f) \, , \\
H_{+}^{(4,4)} &= \frac{1}{2} \Bigg[  \frac{1367 \hat{\mu}_{3}^{2} \hat{\mu}_{4}}{1512}+\frac{16601 \hat{\mu}_{4}}{3465}+\frac{\hat{\mu}_{4} \hat{\epsilon}_{2}^2}{54}-\frac{16 \hat{\mu}_{6}}{495}+\frac{32 \hat{\epsilon}_{5}}{45}  + \bigg( -\frac{1367 \hat{\mu}_{3}^{2} \hat{\mu}_{4}}{216}-\frac{11552 \hat{\mu}_{4}}{693}-\frac{7 \hat{\mu}_{4} \hat{\epsilon}_{2}^2}{54} \, \nonumber \\
&\quad+\frac{16 \hat{\mu}_{6}}{99}-\frac{32 \hat{\epsilon}_{5}}{9} \bigg) \nu + \bigg( \frac{1367 \hat{\mu}_{3}^2 \hat{\mu}_{4}}{126}+\frac{1163 \hat{\mu}_{4}}{231}+\frac{2 \hat{\mu}_{4} \hat{\epsilon}_{2}^2}{9}-\frac{16 \hat{\mu}_{6}}{99}+\frac{32 \hat{\epsilon}_{5}}{9} \bigg) \nu^{2} \, \nonumber \\
&\quad+ \left(1-5 \nu +5 \nu^{2} \right) \left( \frac{16 \hat{\mu}_{6}}{45}-\frac{32 \hat{\epsilon}_{5}}{9} \right) c_{i}^{2} + \Bigg( -\frac{1367 \hat{\mu}_{3}^{2} \hat{\mu}_{4}}{1512}-\frac{16601 \hat{\mu}_{4}}{3465}-\frac{\hat{\mu}_{4} \hat{\epsilon}_{2}^{2}}{54}-\frac{688 \hat{\mu}_{6}}{495}+\frac{128 \hat{\epsilon}_{5}}{45}  \, \nonumber \\ 
&\quad+ \bigg( \frac{1367 \hat{\mu}_{3}^{2} \hat{\mu}_{4}}{216}+\frac{11552 \hat{\mu}_{4}}{693}+\frac{7 \hat{\mu}_{4} \hat{\epsilon}_{2}^2}{54}+\frac{688 \hat{\mu}_{6}}{99}-\frac{128 \hat{\epsilon}_{5}}{9} \bigg) \nu +\bigg( -\frac{1367 \hat{\mu}_{3}^{2} \hat{\mu}_{4}}{126}-\frac{1163 \hat{\mu}_{4}}{231}-\frac{2 \hat{\mu}_{4} \hat{\epsilon}_{2}^2}{9}\, \nonumber \\
&\quad-\frac{688 \hat{\mu}_{6}}{99}+\frac{128 \hat{\epsilon}_{5}}{9} \bigg)\nu^{2}\Bigg) c_{i}^{4} + \frac{16 \hat{\mu}_{6}}{15} \left(1-5 \nu +5 \nu^{2} \right) c_{i}^{6}  \Bigg]  \, \Theta (4 F_{\rm cut}-f) \, , \nonumber \\ 
H_{+}^{(6,4)} &=  \frac{1}{\sqrt{6}} s_{i}^{4} \Bigg[ -\frac{81}{40} \left(1 - 5 \nu + 5 \nu^{2} \right) (1+c_{i}^{2}) \hat{\mu}_{6} \Bigg]  \, \Theta (6 F_{\rm cut}-f) \, ,
\end{align}
\end{subequations}

\begin{subequations}
\label{eq:HcFD}
\begin{align}
H_{\times}^{(2,0)} &= - \frac{1}{\sqrt{2}} \Big[ 2i \, c_{i} \Big] \, \Theta (2 \, F_{\rm cut}-f)\, ,\\
H_{\times}^{(1,1)} &= - s_{i} \, c_{i} \, \delta \bigg[ i \, \Big( \frac{\hat{\mu}_{3}}{12} + \frac{2 \, \hat{\epsilon}_{2}}{3} \Big) \bigg] \, \Theta (F_{\rm cut}-f)\, , \\
H_{\times}^{(3,1)} &= \frac{1}{\sqrt
{3}} \bigg[ \frac{9}{8} \, s_{i} \, \delta \, ( 2i \, c_{i})\, \hat{\mu}_{3} \bigg] \, \Theta (3 \, F_{\rm cut}-f)\, , \\
H_{\times}^{(1,2)} &= s_{i} \, c_{i} \,  \bigg[i \left(\boldsymbol{\chi}_a\cdot\mathbf{\hat{L}}_{\rm N} + \delta \, \boldsymbol{\chi}_s\cdot\mathbf{\hat{L}}_{\rm N}\right) \hat{\epsilon}_{2} \bigg] \, \Theta (F_{\rm cut}-f)\, , \\
H_{\times}^{(2,2)} &= \frac{1}{\sqrt{2}} i \Bigg[ \Bigg(  \frac{3}{2}+\frac{1367 \hat{\mu}_{3}^2}{1008}+\frac{5 \hat{\mu}_{4}}{21}+\frac{\hat{\epsilon}_{2}^2}{36}+\frac{\hat{\epsilon}_{3}}{3}  + \left( \frac{1}{6} -\frac{1367 \hat{\mu}_{3}^2}{252}-\frac{5 \hat{\mu}_{4}}{7}-\frac{\hat{\epsilon}_{2}^2}{9}-\hat{\epsilon}_{3} \right) \nu \Bigg) c_{i}  \nonumber \\
&\quad + \Bigg(  -\frac{\hat{\mu}_{4}}{3} - \hat{\epsilon}_{3} + \left( \hat{\mu}_{4}+3 \hat{\epsilon}_{3}\right) \nu \Bigg) c_{i}^{3} \Bigg] \, \Theta (2 \, F_{\rm cut}-f)\, , \\
H_{\times}^{(4,2)} &= - \frac{1}{2} \, s_{i}^{2} \,  \bigg[ \frac{4}{3} (1-3\nu) \, (2i \, c_{i}) \, \hat{\mu}_{4}  \bigg] \, \Theta (4 \, F_{\rm cut}-f)\, , \\
H_{\times}^{(1,3)} &= i \, s_{i} \, \delta \, \bigg[ \Bigg(\frac{73 \hat{\mu}_{3}}{1008} +\frac{\hat{\mu}_{3} \hat{\epsilon}_{2}^2}{864} +\frac{1367 \hat{\mu}_{3}^2 \hat{\epsilon}_{2}}{3024} + \frac{1367 \hat{\mu}_{3}^3}{24192}+\frac{\hat{\mu}_{5}}{1440}-\frac{50 \hat{\epsilon}_{2}}{63}+\frac{\hat{\epsilon}_{2}^3}{108}+\frac{\hat{\epsilon}_{4}}{35}  \nonumber \\
&\quad+ \left( \frac{173 \hat{\mu}_{3}}{1008}-\frac{\hat{\mu}_{3} \hat{\epsilon}_{2}^2}{216}-\frac{1367 \hat{\mu}_{3}^2 \hat{\epsilon}_{2}}{756}-\frac{1367 \hat{\mu}_{3}^3}{6048}-\frac{\hat{\mu}_{5}}{720}+\frac{19 \hat{\epsilon}_{2}}{42}-\frac{\hat{\epsilon}_{2}^3}{27}-\frac{2 \hat{\epsilon}_{4}}{35} \right)\nu \Bigg)c_{i} \nonumber \\
&\quad+ \Bigg( -\frac{\hat{\mu}_{5}}{480}-\frac{\hat{\epsilon}_{4}}{20}  + \left(\frac{\hat{\mu}_{5}}{240}+\frac{\hat{\epsilon}_{4}}{10} \right) \nu  \Bigg) c_{i}^{3} \bigg] \, \Theta (F_{\rm cut}-f)\, , \\
H_{\times}^{(2,3)} &= \frac{1}{\sqrt{2}} i \, c_{i} \bigg[ -\left(\frac{20}{3}  + \frac{\hat{\epsilon}_{2}^{2}}{12} \right) \delta \left(\boldsymbol{\chi}_a\cdot\mathbf{\hat{L}}_{\rm N} \right) + \Bigg(  -\frac{20}{3} - \frac{\hat{\epsilon}_{2}^{2}}{12}  \nonumber \\
&\quad+ \left( \frac{10}{3}+\frac{\hat{\epsilon}_{2}^2}{3}+\frac{4 \hat{\epsilon}_{3}}{3} -4 c_{i}^2 \, \hat{\epsilon}_{3} \right)\nu \Bigg) \left(\boldsymbol{\chi}_s\cdot\mathbf{\hat{L}}_{\rm N} \right) \bigg] \, \Theta (2 \, F_{\rm cut}-f)\, , \\
H_{\times}^{(3,3)} &= \frac{1}{\sqrt{3}} i \, s_{i} \, \delta \Bigg[ \Bigg(  -\frac{555 \hat{\mu}_{3}}{112} -\frac{\hat{\mu}_{3} \hat{\epsilon}_{2}^2}{32} -\frac{1367 \hat{\mu}_{3}^3}{896}-\frac{27 \hat{\mu}_{5}}{64}  + \left( \frac{153 \hat{\mu}_{3}}{112} + \frac{\hat{\mu}_{3} \hat{\epsilon}_{2}^2}{8} + \frac{1367 \hat{\mu}_{3}^3}{224 }+\frac{27 \hat{\mu}_{5}}{32} \right) \nu \Bigg) c_{i} \, \\
&\quad+  \left(\frac{243 \hat{\mu}_{5}}{320}+\frac{27 \hat{\epsilon}_{4}}{20}\right) \left(1-2\nu\right) \ c_{i}^{3} \Bigg] \, \Theta (3 \, F_{\rm cut}-f)\, , \\
H_{\times}^{(5,3)} &= \frac{1}{\sqrt{5}}  s_{i}^{3} \, \delta \, (1-2\nu) \Bigg[ \frac{625}{384} (2 i \, c_{i}) \hat{\mu}_{5} \Bigg] \, \Theta (5 \, F_{\rm cut}-f)\, , \\
H_{\times}^{(1,4)} &= i \, s_{i} \, c_{i} \Bigg[ \delta \Bigg(i \left(\frac{7 \hat{\mu}_{3}}{60}+\frac{\hat{\epsilon}_{2}}{3}\right) + \left( \frac{\hat{\mu}_{3}}{12} + \frac{2 \hat{\epsilon}_{2}}{3} \right) \pi + i \left( \frac{\hat{\mu}_{3}}{6}  + \frac{4 \hat{\epsilon}_{2}}{3}\right) {\rm ln} \, 2   \Bigg) \, \nonumber \\
&\quad +  \Bigg(  -\frac{2 \hat{\mu}_{3}}{9}-\frac{\hat{\mu}_{3} \hat{\epsilon}_{2}^{2}}{288}-\frac{1367 \hat{\mu}_{3}^{2} \hat{\epsilon}_{2}}{2016}-\frac{247 \hat{\epsilon}_{2}}{252} -\frac{\hat{\epsilon}_{2}^{3}}{24} + \left( -\frac{7 \hat{\mu}_{3}}{24}+\frac{\hat{\mu}_{3} \hat{\epsilon}_{2}^{2}}{72}+\frac{1367 \hat{\mu}_{3}^{2} \hat{\epsilon}_{2}}{504}-\frac{27 \hat{\epsilon}_{2}}{28}+\frac{\hat{\epsilon}_{2}^{3}}{6}+\frac{\hat{\epsilon}_{4}}{14} \right) \nu  \, \nonumber \\
&\quad - \left( \frac{\hat{\epsilon}_{4}}{8} \nu \right)  c_{i}^{2} \Bigg) \delta \left(\boldsymbol{\chi}_s\cdot\mathbf{\hat{L}}_{\rm N} \right) +  \Bigg(  - \frac{2 \hat{\mu}_{3}}{9}-\frac{\hat{\mu}_{3} \hat{\epsilon}_{2}^{2}}{288}-\frac{1367 \hat{\mu}_{3}^{2} \hat{\epsilon}_{2}}{2016}-\frac{247 \hat{\epsilon}_{2}}{252} -\frac{\hat{\epsilon}_{2}^{3}}{24} \,  \nonumber \\
&\quad + \left(\frac{79 \hat{\mu}_{3}}{72}+\frac{\hat{\mu}_{3} \hat{\epsilon}_{2}^{2}}{72}+\frac{1367 \hat{\mu}_{3}^{2} \hat{\epsilon}_{2}}{504}+\frac{325 \hat{\epsilon}_{2}}{252}+\frac{\hat{\epsilon}_{2}^{3}}{6}-\frac{\hat{\epsilon}_{4}}{14} \right) \nu + \left( \frac{\hat{\epsilon}_{4}}{8} \nu \right) c_{i}^{2} \Bigg) \left(\boldsymbol{\chi}_a\cdot\mathbf{\hat{L}}_{\rm N} \right) \Bigg]\, \Theta (F_{\rm cut}-f)\, , \\
H_{\times}^{(2,4)} &= \frac{1}{\sqrt{2}} \, i \, c_{i} \Bigg[ \frac{171}{16} -\frac{419641 \hat{\mu}_{3}^{2}}{84672} -\frac{6835 \hat{\mu}_{3}^{2} \hat{\mu}_{4}}{42336}-\frac{1367 \hat{\mu}_{3}^{2} \hat{\epsilon}_{2}^{2}}{24192}-\frac{1367 \hat{\mu}_{3}^{2} \hat{\epsilon}_{3}}{6048}-\frac{1868689 \hat{\mu}_{3}^{4}}{1354752}-\frac{10049 \hat{\mu}_{4}}{19404}\,  \nonumber \\
&\quad -\frac{5 \hat{\mu}_{4} \hat{\epsilon}_{2}^{2}}{1512} 
+\frac{8965 \hat{\mu}_{4}^2}{3969}-\frac{37 \hat{\mu}_{6}}{5940}+\frac{263 \hat{\epsilon}_{2}^{2}}{3024}-\frac{\hat{\epsilon}_{2}^{2} \hat{\epsilon}_{3}}{216}
-\frac{\hat{\epsilon}_{2}^{4}}{1728}-\frac{437 \hat{\epsilon}_{3}}{3780}+\frac{5 \hat{\epsilon}_{3}^{2}}{63}-\frac{2 \hat{\epsilon}_{5}}{27}  \,  \nonumber \\
&\quad + \bigg( -7+\frac{261073 \hat{\mu}_{3}^{2}}{12096} + \frac{6835 \hat{\mu}_{3}^{2} \hat{\mu}_{4}}{6048}+\frac{1367 \hat{\mu}_{3}^{2} \hat{\epsilon}_{2}^{2}}{3024}+\frac{1367 \hat{\mu}_{3}^{2} \hat{\epsilon}_{3}}{864}+\frac{1868689 \hat{\mu}_{3}^{4}}{169344}+\frac{6250 \hat{\mu}_{4}}{4851} \,  \nonumber \\
&\quad +\frac{5 \hat{\mu}_{4} \hat{\epsilon}_{2}^{2}}{216} - \frac{17930 \hat{\mu}_{4}^{2}}{1323} + \frac{37 \hat{\mu}_{6}}{1188}-\frac{1159 \hat{\epsilon}_{2}^{2}}{3024}+\frac{7 \hat{\epsilon}_{2}^{2} \hat{\epsilon}_{3}}{216}+\frac{\hat{\epsilon}_{2}^{4}}{216}+\frac{80 \hat{\epsilon}_{3}}{189}-\frac{10 \hat{\epsilon}_{3}^{2}}{21}+\frac{10 \hat{\epsilon}_{5}}{27} \bigg)\nu \,  \nonumber \\
&\quad+ \bigg(\frac{19}{144} -\frac{49649 \hat{\mu}_{3}^{2}}{7056} -\frac{6835 \hat{\mu}_{3}^{2} \hat{\mu}_{4}}{3528}-\frac{1367 \hat{\mu}_{3}^{2} \hat{\epsilon}_{2}^{2}}{1512}-\frac{1367 \hat{\mu}_{3}^{2} \hat{\epsilon}_{3}}{504}-\frac{1868689 \hat{\mu}_{3}^{4}}{84672}+\frac{5105 \hat{\mu}_{4}}{6468} \, \nonumber \\
&\quad-\frac{5 \hat{\mu}_{4} \hat{\epsilon}_{2}^{2}}{126}+\frac{8965 \hat{\mu}_{4}^2}{441}-\frac{37 \hat{\mu}_{6}}{1188}+\frac{107 \hat{\epsilon}_{2}^{2}}{756}-\frac{\hat{\epsilon}_{2}^{2} \hat{\epsilon}_{3}}{18}-\frac{\hat{\epsilon}_{2}^{4}}{108}+\frac{31 \hat{\epsilon}_{3}}{756}+\frac{5 \hat{\epsilon}_{3}^{2}}{7}-\frac{10 \hat{\epsilon}_{5}}{27} \bigg)\nu^{2} \, \nonumber \\ 
&\quad+ \Bigg( \frac{1367 \hat{\mu}_{3}^{2} \hat{\mu}_{4}}{6048}+\frac{1367 \hat{\mu}_{3}^{2} \hat{\epsilon}_{3}}{2016}+\frac{10049 \hat{\mu}_{4}}{13860}+\frac{\hat{\mu}_{4} \hat{\epsilon}_{2}^{2}}{216}+\frac{31 \hat{\mu}_{6}}{990}+\frac{\hat{\epsilon}_{2}^{2} \hat{\epsilon}_{3}}{72}+\frac{437 \hat{\epsilon}_{3}}{1260}+\frac{4 \hat{\epsilon}_{5}}{15}  \, \nonumber \\ 
&\quad + \bigg( -\frac{1367 \hat{\mu}_{3}^{2} \hat{\mu}_{4}}{864}-\frac{1367 \hat{\mu}_{3}^{2} \hat{\epsilon}_{3}}{288}-\frac{1250 \hat{\mu}_{4}}{693}-\frac{7 \hat{\mu}_{4} \hat{\epsilon}_{2}^{2}}{216}-\frac{31 \hat{\mu}_{6}}{198}-\frac{7 \hat{\epsilon}_{2}^{2} \hat{\epsilon}_{3}}{72}-\frac{80 \hat{\epsilon}_{3}}{63}-\frac{4 \hat{\epsilon}_{5}}{3} \bigg) \nu \, \nonumber \\
&\quad + \bigg( \frac{1367 \hat{\mu}_{3}^{2} \hat{\mu}_{4}}{504}+\frac{1367 \hat{\mu}_{3}^{2} \hat{\epsilon}_{3}}{168}-\frac{1021 \hat{\mu}_{4}}{924}+\frac{\hat{\mu}_{4} \hat{\epsilon}_{2}^{2}}{18}+\frac{31 \hat{\mu}_{6}}{198}+\frac{\hat{\epsilon}_{2}^{2} \hat{\epsilon}_{3}}{6}-\frac{31 \hat{\epsilon}_{3}}{252}+\frac{4 \hat{\epsilon}_{5}}{3} \bigg) \nu^{2} \Bigg) c_{i}^{2} \, \nonumber \\
&\quad -  \left(\frac{2 \hat{\mu}_{6}}{9}+\frac{\hat{\epsilon}_{5}}{36} \right) \left(1 - 5 \nu +5 \nu^{2} \right) c_{i}^{4} + \left(6+\frac{\hat{\epsilon}_{2}^{2}}{8} \right) \delta \left(\boldsymbol{\chi}_a\cdot\mathbf{\hat{L}}_{\rm N}\right) \left(\boldsymbol{\chi}_s\cdot\mathbf{\hat{L}}_{\rm N}\right)   \, \nonumber \\
&\quad + \left( 3 + \frac{\hat{\epsilon}_{2}^{2}}{16} - 12 \nu  \right) \left(\boldsymbol{\chi}_a\cdot\mathbf{\hat{L}}_{\rm N}\right)^{2} + \left( 3 + \frac{\hat{\epsilon}_{2}^{2}}{16} - \frac{\hat{\epsilon}_{2}^{2}}{4} \nu \right)  \left(\boldsymbol{\chi}_s\cdot\mathbf{\hat{L}}_{\rm N}\right)^{2}  \Bigg]\, \Theta (2 F_{\rm cut}-f)\, ,\\
H_{\times}^{(3,4)} &= \frac{1}{\sqrt{3}} i \, s_{i} \, c_{i} \Bigg[ \delta \bigg( -\frac{189}{20} i + \frac{9 \pi}{4} + \frac{27}{2} i \, {\rm ln}(3/2) \bigg) \hat{\mu}_{3} + \Bigg(  6 \hat{\mu}_{3}+\frac{3 \hat{\mu}_{3} \hat{\epsilon}_{2}^{2}}{32} + \bigg( -\frac{165 \hat{\mu}_{3}}{8}-\frac{3 \hat{\mu}_{3} \hat{\epsilon}_{2}^{2}}{8} \bigg) \nu \, \nonumber \\
&\quad -\frac{27 \hat{\epsilon}_{4}}{8} \, \nu \, c_{i}^{2} \Bigg) \left(\boldsymbol{\chi}_a\cdot\mathbf{\hat{L}}_{\rm N}\right) + \Bigg( 6 \hat{\mu}_{3}+\frac{3 \hat{\mu}_{3} \hat{\epsilon}_{2}^{2}}{32}  - \bigg(\frac{9 \hat{\mu}_{3}}{8}+\frac{3 \hat{\mu}_{3} \hat{\epsilon}_{2}^{2}}{8} \bigg) \nu + \frac{27 \hat{\epsilon}_{4}}{8} c_{i}^{2} \Bigg) \delta \left(\boldsymbol{\chi}_s\cdot\mathbf{\hat{L}}_{\rm N}\right) \Bigg]\, \Theta (3 F_{\rm cut}-f)\, , \\
H_{\times}^{(4,4)} &= \frac{1}{2} i \, c_{i} \Bigg[ \frac{1367 \hat{\mu}_{3}^{2} \hat{\mu}_{4}}{756}+\frac{33202 \hat{\mu}_{4}}{3465}+\frac{\hat{\mu}_{4} \hat{\epsilon}_{2}^2}{27}+\frac{64 \hat{\mu_{6}}}{99}-\frac{16 \hat{\epsilon}_{5}}{45}  + \bigg( -\frac{1367 \hat{\mu}_{3}^{2} \hat{\mu}_{4}}{108}-\frac{23104 \hat{\mu}_{4}}{693}-\frac{320 \hat{\mu}_{6}}{99} \, \nonumber \\
&\quad-\frac{7 \hat{\mu}_{4} \hat{\epsilon}_{2}^{2}}{27}+\frac{16 \hat{\epsilon}_{5}}{9}\bigg) \nu + \bigg( \frac{1367 \hat{\mu}_{3}^{2} \hat{\mu}_{4}}{63}+\frac{2326 \hat{\mu}_{4}}{231}+\frac{4 \hat{\mu}_{4} \hat{\epsilon}_{2}^2}{9} + \frac{320 \hat{\mu_{6}}}{99}-\frac{16 \hat{\epsilon}_{5}}{9} \bigg) \nu^{2} + \Bigg(  -\frac{1367 \hat{\mu}_{3}^2 \hat{\mu}_{4}}{756} \, \nonumber \\
&\quad-\frac{33202 \hat{\mu}_{4}}{3465}-\frac{\hat{\mu}_{4} \hat{\epsilon}_{2}^2}{27}-\frac{1024 \hat{\mu}_{6}}{495}-\frac{64 \hat{\epsilon}_{5}}{45}  + \bigg( \frac{1367 \hat{\mu}_{3}^2 \hat{\mu}_{4}}{108}+\frac{23104 \hat{\mu}_{4}}{693}+\frac{1024 \hat{\mu}_{6}}{99}+\frac{7 \hat{\mu}_{4} \hat{\epsilon}_{2}^2}{27}+\frac{64 \hat{\epsilon}_{5}}{9} \bigg) \nu \, \nonumber \\
&\quad+ \bigg( -\frac{1367 \hat{\mu}_{3}^{2} \hat{\mu}_{4}}{63}-\frac{2326 \hat{\mu}_{4}}{231}-\frac{4 \hat{\mu}_{4} \hat{\epsilon}_{2}^2}{9}-\frac{1024 \hat{\mu}_{6}}{99}-\frac{64 \hat{\epsilon}_{5}}{9} \bigg)\nu^{2} \Bigg) c_{i}^{2} + \Bigg( \frac{64 \hat{\mu}_{6}}{45}+\frac{16 \hat{\epsilon}_{5}}{9} \, \nonumber \\
&\quad+ \bigg( -\frac{64 \hat{\mu}_{6}}{9}-\frac{80 \hat{\epsilon}_{5}}{9} \bigg) \nu + \bigg( \frac{64 \hat{\mu}_{6}}{9}+\frac{80 \hat{\epsilon}_{5}}{9} \bigg) \nu^{2} \Bigg) c_{i}^{4} \Bigg]\, \Theta (4 F_{\rm cut}-f)\, , \nonumber \\ 
H_{\times}^{(6,4)} &=  \frac{1}{\sqrt{6}} s_{i}^{4} \Bigg[ -\frac{81}{40} \left(1 - 5 \nu + 5 \nu^{2} \right) (2 i \, c_{i}) \hat{\mu}_{6} \Bigg]  \, \Theta (6 F_{\rm cut}-f) \, .
\end{align}
\end{subequations}
Since the $k$th harmonic ends at $k$ times the orbital frequency cutoff $F_{\rm cut}$, the step function $\Theta(k \, F_{\rm cut}-f)$ is introduced to make sure each harmonic stops at its proper frequency (see Appendix D of Ref.~\cite{ABFO08} for more details). As a consistency check on the calculation, we confirm the recovery of the corresponding GR expressions of the frequency domain polarizations (see Ref.~\cite{MKAF2016}) in the limit, $\mu_l\rightarrow 1, \epsilon_l\rightarrow 1$.
In this limit, our expressions of $H_{+,\times}^{(k,n)}$ in Eqs.~(\ref{eq:HpFD}) and  (\ref{eq:HcFD}) match with the coefficients of $F_{+,\times}$ in $\mathcal{C}_{k}^{n}$ of Ref.~\cite{MKAF2016}. We provide a complete list  of $H_{+,\times}^{(k,n)}$'s contributing at the 2PN order (including the expressions of $\Psi_{\rm SPA}(f)$, which was derived in Refs.~\cite{Kastha:2018bcr,Kastha:2019}) in the file (\textbf{\emph{supl-mk23.m}}).

We summarize in Table~\ref{tab:multipole_structure} the multipole structure in the PN amplitude based on Eqs.~(\ref{eq:HpFD}) and (\ref{eq:HcFD}). The various multipole moments that contribute to the different PN orders in the amplitude are listed. Note that no multipole parameter appears at the 0PN order in the amplitude. In the time domain, the mass-type quadrupole moment $\mu_2$ contributes to the 0PN amplitude. However, when we compute the Fourier domain waveform using SPA, there appears a factor $(dF/dt)^{1/2}$ (i.e., the square root of the second time derivative of the orbital phase), with F being the orbital frequency, in the denominator of the amplitude [See Eq.~(2.8) of Ref.~\cite{VanDenBroeck:2006qu}, Eq.~(2.11) of Ref.~\cite{VanDenBroeck:2006ar}, and Eqs.~(9) and (10) of Ref.~\cite{ABFO08}]. This factor $(dF/dt)^{1/2}$ also admits a PN series expansion, and the leading-order term in that expansion contains $\mu_{2}$. Therefore, the cancellation between $\mu_{2}$ in the denominator due to the 0PN term in $(dF/dt)^{1/2}$ and $\mu_{2}$ in the numerator due to the 0PN term in the time-domain amplitude is responsible for no multipole parameter to appears at the 0PN order in the frequency domain amplitude.

\begin{center}
	\begin{table}
		\begin{tabular}{||c | c | c |c||} 
			\hline
			PN order  & frequency dependences  & Multipole coefficients  & Harmonics\\ [0.5ex] 
			\hline\hline
			0 PN & $f^{-7/6}$ & & 2nd  \\ 
			\hline
			0.5 PN & $f^{-5/6}$  &  $\mu_2$, $\mu_3$, $\epsilon_2$ & 1st, 3rd\\
			\hline
			1.0 PN & $f^{-1/2}$ &  $\mu_2$, $\mu_3$, $\mu_4$, $\epsilon_2$,  $\epsilon_3$ &  2nd, 4th, $\underline{\rm 1st}$\\
			\hline
			1.5 PN & $f^{-1/6}$ &  $\mu_2$,  $\mu_3$, $\mu_5$,  $\epsilon_2$, $\epsilon_4$, $\underline{\epsilon_3}$ & 1st, 3rd, 5th, $\underline{\rm 2nd}$\\
			\hline
			2.0 PN & $f^{1/6}$ &  $\mu_2$, $\mu_3$, $\mu_4$, $\mu_6$, $\epsilon_2$, $\epsilon_3$, $\epsilon_5$, $\underline{\epsilon_4}$ & 1st, 2nd, 3rd, 4th, 6th\\
			\hline
		\end{tabular}
		\caption{Summary of the multipolar structure in the PN amplitude. The contributions of various multipoles to different PN orders in the amplitude and their frequency dependences are tabulated. We also note down the appearances of different harmonics at each PN orders in the amplitude. The additional multipole parameters and harmonics appearing due to only spin effects are underlined. Following the definitions introduced in Ref.~\cite{Kastha:2018bcr}, $\mu_l$ refer to contribution from the mass-type multipole moments and $\epsilon_{l}$ refer to the contribution from the current-type multipole moments.}
		\label{tab:multipole_structure}
	\end{table}
\end{center}

\section{Conclusion}\label{sec:conclu}
A parametrized multipolar gravitational wave amplitude is obtained in this paper. The parametrization is adopted following the previous works~\cite{Kastha:2018bcr, Kastha:2019} to trace the contributions from the radiative multipole moments in the gravitational wave amplitude. This generic multipolar parametrization of the amplitude together with the phase reported in Refs.~\cite{Kastha:2018bcr, Kastha:2019} not only facilitates a unique test of the multipolar structure of compact binary inspiral but the amplitude corrections obtained here are expected to put even more stringent bounds on the multipole parameters.

Here we provide the closed-form expression for the multipolar amplitude of the gravitational waveform at 2PN in both the time and frequency domains. This ready-to-use amplitude-corrected parametrized multipolar waveform consists of amplitude accuracy up to 2PN with phase being accurate upto 3.5PN, which can now be used to compute the projected accuracy on the various multipole moments using GW events in different networks of ground-based detectors and the space-based detector LISA. This will be discussed in a companion paper \cite{Mahapatra:2023uwd}.

Recently, Ref.~\cite{Blanchet:2023bwj} reported the expression of 4.5PN phasing for nonspinning compact binaries using the ingredients developed in Refs.~\cite{Blumlein:2020pog,Bini:2020nsb,Blumlein:2021txe,Larrouturou:2021dma,Larrouturou:2021gqo,Blanchet:2022vsm,Henry:2021cek,Trestini:2022tot,Trestini:2023wwg,Blanchet:2023sbv,Marchand:2020fpt}. The 3PN accurate polarizations for nonspinning binaries was derived in Refs.~\cite{ABIQ04,BFIS08}. The tail-type spin-orbit corrections entering the 3PN amplitude~\cite{BBF2011} and the spin-orbit effects at the 4PN order~\cite{MKAF2016} are also available. Using these ingredients, we plan to compute parametrized multipolar waveform to higher PN order in the phase and the amplitude in future. 
Similarly, constructing a parametrized multipolar waveform model that accounts for the effects of orbital eccentricity and the effects of spin precession is also planned. Reference~\cite{Mahapatra:2023hqq} proposed a new parametrization to probe the different PN components in radiative mass-type octupole moment (See Eq.~(4) of Ref.~\cite{Mahapatra:2023hqq}). Such parametrizations can be adopted for other radiative multipole moments to keep track of the contributions from different PN orders in radiative moments to the GW phasing and amplitude. Computing the multipolar waveform with such parametrizations is also among the plans we want to pursue.

\acknowledgements
We thank K. G. Arun for fruitful discussions and comments on the manuscript. We also acknowledge A. Gupta and B. S. Sathyaprakash for the invaluable discussions and comments on the manuscript. P.M. thanks A. Laddha for the useful discussions. 
P.M. acknowledges the support of the Core Research Grant No. CRG/2021/004565 of the Science and Engineering Research Board of India and a grant from the Infosys Foundation.
S.K. acknowledges support from the Villum Investigator program supported by the VILLUM Foundation (Grant No. VIL37766) and the DNRF Chair program (Grant No. DNRF162) by the Danish National Research Foundation. This project has received funding from the European Union's Horizon 2020 research and innovation programme under the Marie Sklodowska-Curie Grant Agreement No. 101131233.


\bibliography{ref-list}

\end{document}